\documentclass[11pt,a4paper]{article}
\usepackage[T1]{fontenc}
\usepackage[cp1250]{inputenc}
\usepackage{graphicx}
\usepackage{wrapfig}
\usepackage{amsmath}
\usepackage{bm}
\usepackage{amssymb}
\usepackage{cancel}
\usepackage{xcolor}

\usepackage{changes}

\makeatletter
\usepackage{babel}
\makeatother

\usepackage{changes}

\def\m{{\mbox{\scriptsize m}}}
\def\pl{{\mbox{\scriptsize pl}}}
\def\i{{\mbox{\scriptsize I}}}
\def\ii{{\mbox{\scriptsize II}}}
\def\e{{\mbox{\scriptsize e}}}
\def\h{{\mbox{\scriptsize h}}}
\def\a{{\mbox{\scriptsize a}}}
\def\br{{(\pm)}} \def\bp{{(+)}}
\def\bm{{(-)}}
\def\ph{{\mbox{\scriptsize ph}}}

\def\c{{\mbox{\scriptsize c}}}
\def\r{{\mbox{\scriptsize ref}}}
\def\p{{\mbox{\scriptsize phys}}}
\def\o{{\mbox{\scriptsize obs}}}
\def\ct{{\mbox{\scriptsize crit}}}
\def\re{{\mbox{\scriptsize rev}}}
\def\in{{\mbox{\scriptsize i}}}
\def\l{{\mbox{\scriptsize lim}}}
\def\nr{{\mbox{\scriptsize NR}}}

\begin{document}

\renewcommand{\figurename}{Fig.}

\centerline {\LARGE \bf Radiation in the black hole--plasma
system:}

\centerline {\LARGE \bf propagation in equatorial plane}

\vskip 5mm \centerline{\large Vladim\' ir Balek}

\centerline{\it Department of Theoretical Physics, Comenius
University, Bratislava}

\vskip 2mm \centerline{\large Barbora Bezd\v ekov\' a, Ji\v r\' i
Bi\v c\' ak}

\centerline{\it Institute of Theoretical Physics, Charles
University, Prague}

\vskip 1cm

\large Effect of cold plasma on the form of rays propagating in
the equatorial plane of a rotating black hole is investigated. Two
kinds of regions in the radius--impact parameter plane allowed for
the rays are constructed: for radiation with a given frequency at
infinity and for radiation with a given ``telescope frequency''
seen by a local observer. The form of allowed regions for locally
nonrotating observers as well as observers falling freely from
infinity is established. The allowed regions contain rays which
directly reach the horizon, or there exists a ``neck'' connecting
the forbidden regions such that the rays coming from infinity
cannot reach the horizon. In case we considered a set of observers
at various radii instead of the neck we find two different regions
-- from one the rays reach the horizon and not infinity and from
the other one they reach infinity, but not the horizon. The
results are analyzed by analytical methods and illustrated by
figures constructed numerically.

\vskip 1.5cm

{\Large \bf 1. Introduction}

\vskip 5mm It is widely acknowledged that in 1911 Albert Einstein,
in his best known paper written during his Prague stay, addressed
the question of the influence of gravity on light (``\"{U}ber den
Einflu\ss{} der Schwerkraft auf die Ausbreitung des Lichtes'') and
suggested that the bending of rays moving close to the Sun should
be observable. Less is known that Einstein had derived the lens
equation, predicated the double images and the magnifications in
the year 1912, as documented in his notebook written during his
visit of Berlin from Prague \cite{Renn}. As a consequence of many
significant developments in both theory and observations, the
effects of gravity on the light propagation have become subject of
numerous studies. For very comprehensive reviews on gravitational
lensing, see, for example,
\cite{Schneider,Perlick00,Wambsganss,Blandford}.

\vskip 2mm In the following we do not address directly
gravitational lensing, rather, we turn to more general question
about the existence of allowed regions for rays propagating in the
gravitational field of a rotating (Kerr) black hole which is
surrounded by a cold plasma.

\vskip 2mm Let us first mention works on light rays in the fields
of black holes in vacuum. The pioneering study of the allowed
regions for both photons and free massive particles around a Kerr
black hole in the equatorial plane were done many years ago by De
Felice \cite{DeFelice}. He determined and analyzed the turning
points in terms of impact parameter, and constructed also diagrams
corresponding to various particle energies and black hole angular
momenta. Allowed regions in lensing problems around a Kerr black
hole in vacuum were more recently discussed, e.g., by Vazquez and
Esteban \cite{Vazquez} and Gralla and Lupsasca \cite{Gralla}. The
values of impact (``critical'') parameter for which circular light
rays exist were derived by Iyer and Hansen \cite{Iyer}. Their
paper also includes the study of ray deflection angles, proving
that there is the light bending asymmetry due to hole's rotation.

\vskip 2mm For related works on the photon escape cones from black
hole's vicinity, see \cite{Bardeen,Semerak,Schee} and a new
comprehensive review by Ogasawara and Igata \cite{Ogasawara}.

\vskip 2mm Only recently the light ray trajectories in Kerr
spacetime were studied assuming also the presence of a medium in
the spacetime (see, e.g., \cite{Perlick17,Fathi,Chowdhuri}  and
references therein).

\vskip 2mm In the present paper, we study the allowed regions for
rays in cold plasma in the vicinity of a Kerr black hole. Our
basic parameters to characterize the rays are the impact parameter
at infinity and the frequency of radiation. In case of the
frequency we distinguish two physically distinct situations:
either the frequency $\omega$ is fixed at infinity, or we consider
a given ``telescope frequency'' $\omega_0$ which is fixed by the
requirement that it is measured by a given observer located
outside the horizon.

\vskip 2mm Unlike in the study of De Felice \cite{DeFelice}, we
investigate the turning points for rays around a Kerr black hole
surrounded by a refractive and dispersive medium in geometrical
optics approximation by using the elegant Hamiltonian formalism
 by Synge \cite{synge}. This formalism was employed in
developing the radiation transfer theory in refractive and
dispersive media in curved spacetimes by Bi\v{c}\'{a}k and Hadrava
\cite{Bicak}, for example, and it has been used recently in a
number of studies to describe ray behavior in refractive and
dispersive media in the vicinity of black holes. The focus is
usually on gravitational lensing, on black hole shadow etc. Let us
mention just few examples. In the series of papers by
Bisnovatyi-Kogan and Tsupko, e.g.,
\cite{BKT09,BKT13,BKT19,Tsupko}, the authors found the deflection
angle formulas in dispersive media around a Schwarzschild black
hole. The form of the shadow of a Kerr black hole surrounded by
plasma was derived by Perlick and Tsupko \cite{Perlick17} and
further investigated by Chowdhuri and Bhattacharyya
\cite{Chowdhuri}. Fathi et al. \cite{Fathi} used the separability
condition for the Hamilton-Jacobi equation from \cite{Perlick17},
assumed some simple ansatz for the radial and longitudinal
dependence of plasma frequency, and calculated the light rays
analytically in terms of elliptic integrals. More generally,
separability of the Hamilton-Jacobi equation and shadow in axially
symmetric spacetimes with plasma was recently analyzed in
\cite{Bezdekova}.

\vskip 2mm The dispersion of light rays around a Kerr black hole
caused by the presence of a plasma with a power distributed
density was studied by Kimpson et al. \cite{Kimpson}.
S\'{a}ren\'{y} and Balek \cite{Sareny} investigated light rays in
several models of plasma in the vicinity of a Kerr black hole.

\vskip 2mm In our work, we employ the distribution of nonsingular
isothermal sphere \cite{Hinshaw} for plasma. The distribution of a
(non)singular isothermal sphere lies somewhere between exponential
and power-law distributions used as density models for spiral and
elliptical galaxies, respectively \cite{ErMao}. It is frequently
used in the description of gravitational lensing by galaxies or
clusters of galaxies \cite{BKT}. In such studies, the
inhomogeneity parameter in the plasma distribution is by many
orders of magnitude greater than the size of a typical black hole
in the center of galaxy. We will be interested in the effect of
plasma on the propagation of radiation in close proximity of the
black hole, therefore we will suppose the parameter to be of the
same order as the radius of the horizon.

\vskip 2mm In the following section the Hamiltonian method for
rays in a curved spacetime with a (non-gravitating) medium
described by a spacetime dependent and frequency dependent index
of refraction $n$ is introduced. The medium is then considered to
be a cold non-magnetic plasma characterized by the electron
density $N_e$ and plasma frequency $\omega_{pl}$. In order to
determine regions to which the light rays can propagate, we have
to satisfy conditions that (i) the Hamilton equations are
satisfied, and (ii) the constraint of the vanishing Hamiltonian is
fulfilled. The latter condition guarantees that the index of
refraction remains real along any given ray. The conditions are
specified in the Kerr metric. The frequency of the radiation as
seen by a general observer is expressed, in particular for the
observers locally non-rotating (LNRF) and for observers freely
falling from the rest from infinity.

\vskip 2mm In Section 3 allowed regions for the radiation with the
fixed frequency as measured at infinity are analyzed in detail in
analytic terms, and specific examples are constructed numerically.
First the vacuum case is considered, then plasma is added. As
could be anticipated, the forbidden regions are larger than in
vacuum. This is well seen in the diagrams showing the radial
coordinate a ray can reach when having given impact parameter and
frequency at infinity. Frequency can, of course, be negative as
seen at infinity if a ray is close to the horizon and cannot reach
infinity. We pay a particular attention to the formation of
``necks'' in the radius and impact parameters plane which connect
forbidden regions.

\vskip 2mm In Section 4 a detailed analysis is devoted to the
illustration of the allowed regions when a ray should reach a
given LNRF observer with a given frequency $\omega_0$. In
Appendices A--D we give more detailed discussion of the formation
of the neck in both cases when $\omega$ or $\omega_0$ are fixed,
we investigate the allowed regions far/near to infinity, and
provide the analysis of the allowed regions when LNRF observer's
position and the impact parameter are given.

\vskip 2mm In Section 5 the allowed regions are analyzed for
observers freely falling from the rest at infinity. Again, both
analytical and graphical and numerical results are presented. We
give detailed explanations of the results by combining both
analytical considerations and number of figures constructed
numerically. One can concentrate primarily on the introductions to
each section and on the figure captions if one does not wish to
follow detailed ``analytical'' argumentations.

\vskip 2mm We use the system of units $c = G = 1$ and the
signature of the metric $(- + + +)$.

\vskip 5mm {\Large \bf 2. Allowed regions for radiation in the
medium in the Kerr space\-time}

\vskip 5mm Consider an electromagnetic wave propagating in a
refractive and dispersive medium with refractive index $n$ in flat
spacetime. Denote the frequency and the wave vector of the wave in
the rest frame of the medium by $\omega_\m$ and $\bf k_\m$,
respectively. The dispersion relation in this frame, $|{\bf k}_\m|
= n\omega_\m$, can be rewritten in a Lorentz invariant form as
\begin{equation}
k^2 - (n^2 - 1)\omega_\m^2 = 0, \quad \omega_\m = - k \cdot u_\m,
\label{eq:disp}
\end{equation}
where $k^\mu$ is the wave 4-vector and $u_\m^\mu$ is the
4-velocity of the medium. The equation stays valid, with the
replacement $\eta_{\mu \nu} \to g_{\mu \nu}$, for radiation
propagating in an inhomogene\-ous medium around a gravitating body
in a curved spacetime, provided the wavelength of the radiation
$\lambda$ is much smaller than the typical scale $l$ on which
gravitational field and the properties of the medium are varying
({\it geometrical optics approximation}). In this limit, radiation
propagates along {\it rays}, the paths in the spacetime given by
the Hamiltonian of the form \cite{synge}
\begin{equation}
H = -\tfrac 12 \big[k^2 - (n^2 - 1)\omega_\m^2\big]. \label{eq:H}
\end{equation}
The Hamiltonian is subject to the constraint $H = 0$. Rays can be
viewed as worldlines of {\it light signals}, configurations of
electromagnetic field localized in a volume extended over many
$\lambda$'s, but still small in comparison with $l$.

\vskip 2mm The refractive index of a cold non-magnetic plasma can
be expressed in terms of plasma frequency $\omega_\pl$, which in
turn is given by electron number density $N_\e$,
\begin{equation}
n^2 = 1 - \frac {\omega_\pl^2}{\omega_\m^2}, \quad \omega_\pl^2 =
\frac {e^2}{\epsilon_0 m_\e} N_\e. \label{eq:npl}
\end{equation}
The resulting dispersion relation is
\begin{equation}
k^2 + \omega_\pl^2 = 0. \label{eq:disp1}
\end{equation}
The Hamiltonian, $H = -\frac 12 (k^2 + \omega_\pl^2)$, corresponds
to a ``particle'' with variable mass $m \propto \omega_\pl$. If we
identify momentum $p_\mu$ canonically conjugated with spacetime
coordinates $x^\mu$ with the covariant components of the wave
4-vector, $p_\mu = k_\mu$, the 4-velocity of the light signal will
be $\dot x^\mu = k^\mu$, where the dot denotes differentiation
with respect to the parameter of the ray. Note, however, that the
{\it physical} momentum differs from $k^\mu$; in particular, for a
single photon we have $p_\p^\mu = \hbar k^\mu$ and $m_\p =
\hbar\omega_\pl$.

\vskip 2mm Since the Hamiltonian for rays in cold plasma does not
contain the frequency $\omega_\m = - k \cdot u_\m$, the
propagation of light signals is not affected by the motion of
plasma. Still, frequency $\omega_\m$ enters the condition
determining where the radiation can in principle be observed: the
rays can penetrate only regions in which $n$ is real, i.e. the
inequality $\omega_\m \ge \omega_\pl$ is satisfied. We would need
to add this condition to the theory if the form of the rays would
be given just by Hamilton equations. However, the theory contains,
apart from Hamilton equations, an additional constraint that the
Hamiltonian, of which we know from the equations only that it is
conserved, is to be zero. As a consequence, the light signals have
$\omega_\m = ({\bf k}_\m^2 + \omega_\pl^2)^{1/2}$, so the
condition $\omega_\m \ge \omega_\pl$ is fulfilled automatically.

\vskip 2mm Consider Kerr metric in Boyer-Lindquist coordinates
$(x^0, x^1, x^2, x^3) = (t, r, \theta, \phi)$ and suppose the
space outside the horizon is filled with plasma with the same
symmetries as are those of the metric. The system has mirror
symmetry w.r.t. the equatorial plane ($\theta = \pi/2$),
which makes it possible for rays to lie entirely in that plane.
Metric tensor restricted to the equatorial plane is block
diagonal, $g_{\mu \nu} = \displaystyle
  \bigg( \mbox{\hskip -2mm}
  \left. \begin{array} {cc} g_{A B} & 0\\ 0 & g_{11}\\
  \end{array} \right.\bigg)$,
where $A, B = 0, 3$, so $k^2 = g^{A B} k_A k_B + g_{11}\dot
r^2$ and we can rewrite the dispersion relation (\ref{eq:disp1})
as
\begin{eqnarray}
\dot r^2 = -g^{11} (g^{A B} k_A k_B + \omega_\pl^2).
\label{eq:dr2}
\end{eqnarray}
The covariant $t$- and $\phi$-components of the wave 4-vector can
be written as $k_0 = -\omega$ and $k_3 = \omega b$, where $\omega$
and $b$ are frequency and impact parameter, respectively, measured
by distant observers (assuming the rays can reach them). For the
given values of $\omega$ and $b$, the light signal can reach the
given radius $r$ only if the r.h.s. of (\ref{eq:dr2}) is
non-negative, the zero value corresponding either to a turning
point or to a circular orbit.

\vskip 2mm Kerr metric depends on two parameters: the black hole
mass $M$ and the Kerr parameter $a$ (angular momentum per unit
mass). If we put $M = 1$, denote $u = r^{-1}$ and introduce
functions $f = 1 - 2u$, ${\cal A} = 1 + a^2 u^2 + 2a^2 u^3$ and
${\cal D} = 1 - 2u + a^2 u^2$, the metric in the equatorial plane
becomes
\begin{eqnarray*}
ds^2 = -fdt^2 - 4ar^{-1}dtd\phi + {\cal A}r^2d\phi^2 + {\cal
D}^{-1}dr^2.
\end{eqnarray*}
Note that functions $f$, $\cal A$ and $\cal D$ satisfy the
identity $f{\cal A} + 4a^2u^4 = {\cal D}$. By inserting for
$g^{AB}$, $k_0$ and $k_3$, we obtain (denoting by
superscript ``$\circ$'' the vacuum case)
\begin{eqnarray*}
g^{A B} k_A k_B = -\omega^2 \frac {\mathring{F}}{{\cal D}r^2},
\quad \mathring{F} = {\cal A}r^2 - 4ar^{-1}b - fb^2.
 \label{eq:k2}
\end{eqnarray*}
If we plug this into equation (\ref{eq:dr2}) and use $g^{11} =
\cal D$, we find
\begin{eqnarray}
\dot r^2 = \omega^2 Fr^{-2}, \quad \text{where in plasma}\ F =
\mathring{F} - {\cal D}r^2 \frac{\omega_\pl^2}{\omega^2}.
 \label{eq:dfF}
\end{eqnarray}
Light signals can propagate only in the regions in the $(r,
b)$-plane with $F \ge 0$; they have turning points ($\dot r=0$) at
$F = 0$ and revolve around the black hole on a circular orbit if,
in addition to $F = 0$, $F$ also satisfies $F' = 0$ (prime
denoting differentiation with respect to $r$).

\vskip 2mm We are interested in regions in the $(r, b)$-plane
accessible to rays in the equatorial plane of Kerr black hole
surrounded by plasma with the given electron number density
$N_\e$. We consider two situations: (i) the light signal
propagating along the ray has a fixed frequency $\omega$ at
infinity, or (ii) it has a fixed ``telescope frequency''
$\omega_0$ seen by an observer with given 4-velocity $u_\o^\mu$ at
the given radius $r_0$. In the latter case the spectrum of the
light arriving at the observer, be it from distant sources (stars)
or from a source close to the horizon, is supposed to spread
across all frequencies $\omega$, so that among the components of
the light there is always one whose frequency in the observer's
frame is $\omega_0$. For plasma, we will employ the model of a
{\it nonsingular isothermal sphere} used, for example, in
\cite{BKT,Liu}, in which $N_\e \propto (r^2 + r_\c^2)^{-1}$. In
the problem with fixed ``telescope frequency'' we will consider
first an observer at rest with respect to a given {\it locally
nonrotating frame} (LNRF), and then an observer falling freely
from rest at infinity.

\vskip 2mm The 4-velocity of a particle moving in the equatorial
plane is $u^\mu = \Gamma(1, v, 0, \Omega)$, where $v$ and $\Omega$
are radial and angular coordinate velocities, $v = dr/dt$ and
$\Omega = d\phi/dt$; $\Gamma= dt/d\tau$ is the conversion factor
between the time of distant observers $t$ and the proper time
$\tau$ of the particle. Let $\Omega$ and $\Gamma$  be the angular
velocity and the Lorentz factor of LNRF, $\Omega = 2au^3/{\cal
A}$, $\Gamma = ({\cal A}/{\cal D})^{1/2}$, and $\hat v$ and $\hat
\Gamma$ be the (coordinate) radial velocity and the Lorentz factor
of a particle falling freely from the rest at infinity, $\hat v =
-\ {\cal D}\sqrt{\alpha}/{\cal A}$, where $\alpha = {\cal A} -
{\cal D} = 2u(1 + a^2 u^2)$, and $\hat \Gamma = {\cal A}/{\cal
D}$. (This follows from separated geodesic equations --  see
\cite{BiSt} for a detailed discussion, including the integration
in this case.) Using this notation, we have $v_\o = 0$, $\Omega_\o
= \Omega_0$ and $\Gamma_\o = \Gamma_0$ (index 0 refers to the
radius $r_0$) for a locally nonrotating observer, and $v_\o = \hat
v_0$, $\Omega_\o = \Omega_0$, and $\Gamma_\o = \hat \Gamma_0$ for
a freely falling observer.

\vskip 2mm The frequency seen by an observer with the 4-velocity
$u_\o^\mu$ is $\omega_0 = - k \cdot u_\o = \omega \Gamma_\o(1 -
{\cal D}_0^{-1} v_\o \eta_0 - \Omega_\o b)$, where $\eta$ is the
rescaled radial velocity of light, $\eta = \dot r_\ph/\omega = \pm
\sqrt{F}u$ (index ``ph'' stands for ``photon''). For
the locally nonrotating observer this yields
\begin{eqnarray}
\omega_0 = \omega \Gamma_0(1 - \Omega_0 b),
 \label{eq:om0}
\end{eqnarray}
and for the freely falling observer ($\xi = {\cal D}^{-1} \hat v =
-\sqrt{\alpha}/{\cal A}$)
\begin{eqnarray}
\omega_0 = \omega \hat \Gamma_0(1 - \xi_0 \eta_0 - \Omega_0 b).
 \label{eq:om0f}
\end{eqnarray}

\vskip 3mm {\Large \bf 3. Radiation with fixed $\omega$}

\vskip 5mm Consider first radiation in vacuum. The boundaries of
allowed regions are given by the solutions to the quadratic
equation $\mathring{F} = 0$, which can be expressed in two
equivalent forms,
\begin{eqnarray}
\mathring{b}_\pm = \frac 1f (-2ar^{-1} \pm \sqrt{{\cal D}}r) =
\frac {{\cal A}r^2}{2ar^{-1} \pm \sqrt{\cal D}r}. \label{eq:bpm0}
\end{eqnarray}
Functions $\mathring{b}_\pm$ coincide at the horizon,
$\mathring{b}_\pm = b_\h = \Omega_\h^{-1}$ at $r = r_\h = 1 +
\sqrt{1 - a^2}$ (the larger root
\begin{wrapfigure}[19]{l}{7cm}
\vskip -2.5mm
\includegraphics [height=0.32\textheight]{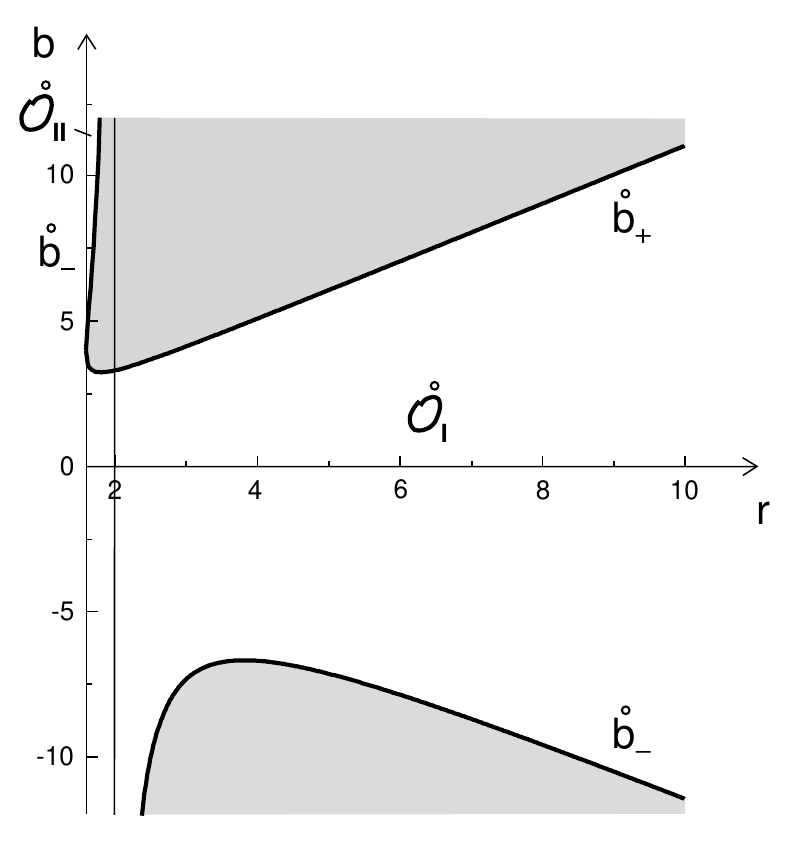}
\centerline{\parbox {6cm}{\caption{\small Allowed regions for rays
in vacuum. Kerr parameter is $a = 0.8$ and radius of the horizon
is $r_\h = 1.6$. Shaded areas bounded by the lines
$\mathring{b}_+$ and $\mathring{b}_-$ (``corners'') are forbidden,
free areas $\mathring{\cal O}_\i$ and $\mathring{\cal O}_\ii$
outside them are al\-lowed. The thin vertical line at $r = 2$ is
the asymptote of the inner boundary of forbidden regions.
\label{fig:allV}}}}
\end{wrapfigure}
of ${\cal D} = 0$), and approach asymptotes at the angles $\pm
45^\circ$ to the $r$ axis far from the horizon, $\mathring{b}_\pm
\sim \pm r$ for $r \gg 1$. From the first expression in
(\ref{eq:bpm0}) we can see that the $\mathring{b}_-$ root is
falling to $-\infty$ and rising to $+\infty$ as we approach the
radius $r = 2$ (static limit) from the left and right,
respectively. From the second expression in (\ref{eq:bpm0}), on
the other hand, it is seen that the $\mathring{b}_+$ root stays
finite at $r = 2$. As seen in Fig. 1, there are two distinct
regions in the $(r, b)$-diagram forbidden for the rays (regions
where $\mathring{F} < 0$): the ``upper corner'' above
$\mathring{b}_+$ at $r > 2$ and between $\mathring{b}_+$ and
$\mathring{b}_-$ at $r < 2$, and the ``lower corner'' under
$\mathring{b}_-$ at $r > 2$. Outside the corners there are two
regions accessible for rays, region $\mathring{\cal O}_\i$ between
the corners at $r > 2$ and below the upper corner at $r < 2$, and
region $\mathring{\cal O}_\ii$ left to the upper corner at $r <
2$. Photon orbits are located at the lowest point of the upper
corner (corotating orbit) and at the highest point of the lower
corner (counter-rotating orbit).

\vskip 2mm Frequency of radiation seen by an observer with $v_\o =
0$ is $\omega_\o = \omega \Gamma_\o(1 - \Omega_\o b)$. This
frequency is necessarily $> 0$, hence we can determine the sign of
$\omega$ at any given point in the $(r, b)$-plane by analyzing the
interval of possible values of the angular velocity $\Omega_\o$.
Denote by $\Omega_\pm$ angular velocities of co- and
counter-rotating light signals in vacuum, kept on a path close to
a circle with an arbitrary given radius by a system of mirrors
tangential to the circle and placed at equal distances from each
other. The signals have, in the limit when the number of mirrors
goes to infinity, $b = \mathring{b}_\pm$, and their 4-velocity
squared is $u_\ph^2 = -\omega \dot t_\ph + \kappa \dot \phi_\ph
\propto -1 + b\Omega_\ph = 0$. Thus, $\mathring{b}_\pm$ are just
inverses to $\Omega_\pm$; $\mathring{b}_\pm = \Omega_\pm^{-1}$.
The value of $\Omega_\o$ falls between $\Omega_{-0}$ and
$\Omega_{+0}$, therefore, $b/\mathring{b}_{-0} < \Omega_\o b <
b/\mathring{b}_{+0}$ for $b > 0$ and $\Omega_\o b <
b/\mathring{b}_{-0}$ for $b < 0$. As a result, we have different
signs of $\omega$ in different allowed regions: light signals have
$\omega > 0$ in region $\mathring{\cal O}_\i$ and $\omega < 0$ in
region $\mathring{\cal O}_\ii$. The radiation with $\omega < 0$ is
a special case of a ``particle'' with a negative energy -- a
necessary ingredient of the Penrose process (see, e.g.,
\cite{MTW}, \S 33.7).

\vskip 2mm Now let us add plasma to the picture. Consider
radiation with a fixed frequency $\omega$ at infinity, positive in
region $\mathring{\cal O}_\i$ and negative in region
$\mathring{\cal O}_\ii$. The function defining the boundaries of
the allowed regions is now shifted, $F = \mathring{F} - {\cal
D}r^2 \zeta^2$, where $\zeta = \omega_\pl/\omega$. Equation $F =
0$ is again quadratic, and its solutions $b_\pm$ can be written in
the form of the first expression in (\ref{eq:bpm0}) with ${\cal D}
\to (1 - f\zeta^2){\cal D}$ (equation (\ref{eq:bpm})). We can see
that functions $b_\pm$ have the same asymptotes at $r \gg 1$ as
functions $\mathring{b}_\pm$, $b_\pm \sim \pm r$, and that the
asymptote of function $b_-$, with $b_-$ falling to $-\infty$ to
the right of it and rising to $+\infty$ to the left of it, lies at
$r = 2$ just like that of function $\mathring{b}_-$. However, the
similarity between functions $\mathring{b}_-$ and $b_\pm$ ends
there: $b_\pm$, when compared to $\mathring{b}_\pm$, are pulled
towards each other or entirely vanish at $r > 2$, and are pushed
away from each other at $r < 2$. As a result, both forbidden
regions expand, and there may even appear a neck between them, so
that no ray, regardless of its impact parameter, can pass from
black hole to distant observers or {\it vice versa}. The formation
of the neck is discussed in Appendix A.

\vskip 2mm As the forbidden regions expand, allowed regions
shrink: instead of regions $\mathring{\cal O}_\i$ and
$\mathring{\cal O}_\ii$ in vacuum we have smaller regions ${\cal
O}_\i$ and ${\cal O}_\ii$ contained within them. The former region
is divided into two parts in the presence of a neck, ${\cal O}_{\i
A}$ on the lower left of the neck and ${\cal O}_{\i B}$ on its
right. Since ${\cal O}_\i$ and ${\cal O}_\ii$ lie inside
$\mathring{\cal O}_\i$ and $\mathring{\cal O}_\ii$, we have
$\omega > 0$ in the former region and $\omega < 0$ in the latter
one.

\vskip 2mm Function $\zeta$ is proportional to $\omega_\pl$, hence
to $\sqrt{N}_\e$, so, for our choice of plasma distribution, to
$(r^2 + r_\c^2)^{-1/2}$. To fix the coefficient of
proportionality, we choose some radius $r_\r$ close to $r_\h$ and
some frequency ratio $q = \omega_{\pl,\r}/|\omega|$, and write
$\zeta = \pm q\sqrt{\cal R}$, where ${\cal R} = N_\e/N_{\e,\r} =
(r_\r^2 + r_\c^2)/(r^2 + r_\c^2)$; the signs $+$ and $-$ refer to
region ${\cal O}_\i$ and ${\cal O}_\ii$, respectively.

\begin{figure}[ht]
\centerline{\includegraphics[height=0.32\textheight]{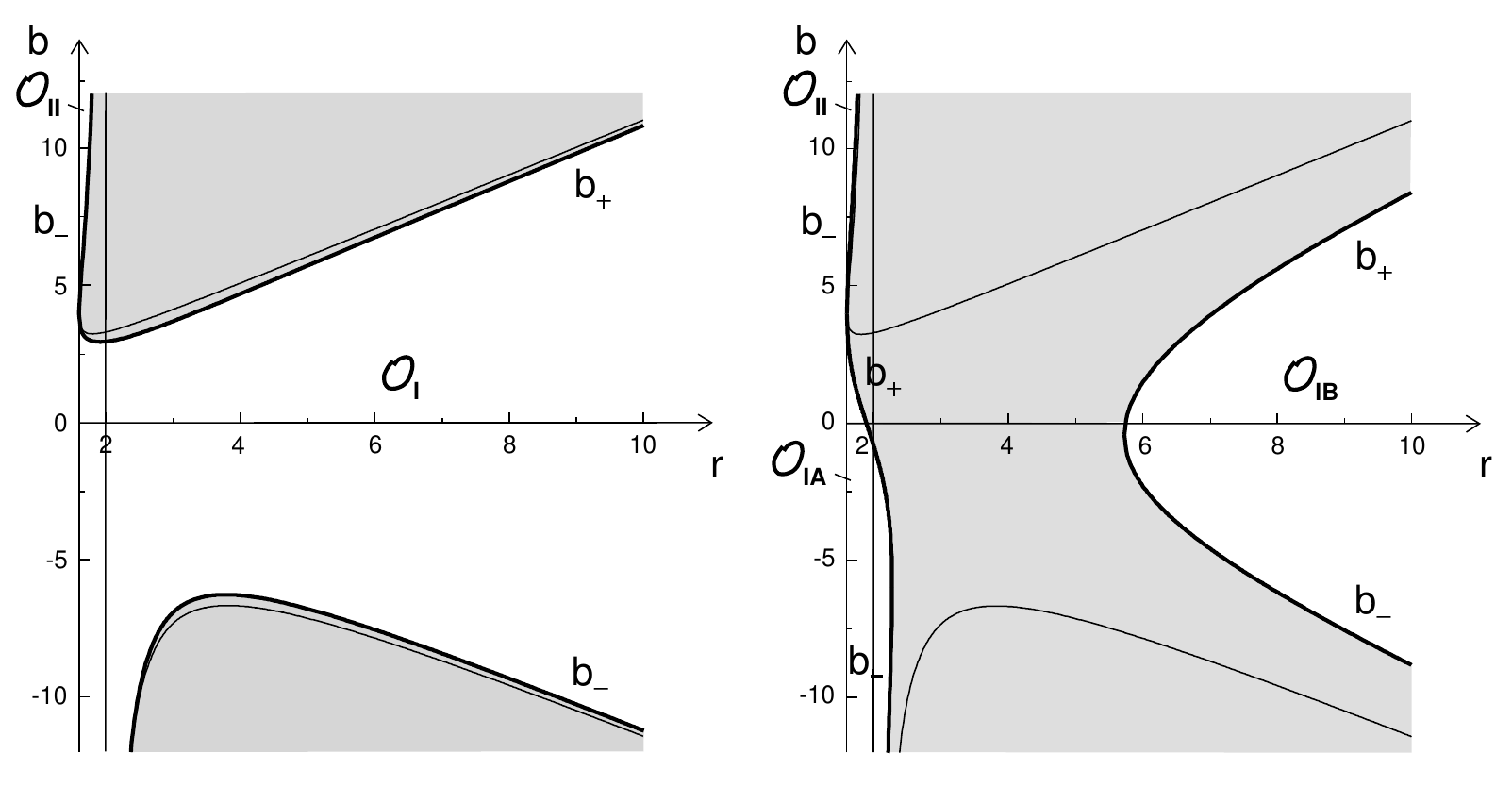}}
\centerline{\parbox {16.2cm}{\caption{\small Allowed regions for
rays in plasma with fixed frequency $\omega$ at infinity. In
addition to $a = 0.8$ we have $r_\c = 1$, $r_\r = 1.8$ (the
reference circle lies halfway between the horizon and the static
limit) and $q = 1$, 3.5 in the left and right panel, respectively.
Shaded areas are forbidden, free areas outside them, regions
${\cal O}_\i$ and ${\cal O}_\ii$ in the left panel and regions
${\cal O}_{\i A}$, ${\cal O}_{\i B}$ and ${\cal O}_\ii$ in the
right panel, are allowed. For comparison, the boundaries of
forbidden regions in vacuum are shown, too, depicted by lighter
lines inside the shaded regions. The vertical line, located at the
static limit just like in vacuum, is the asymptote of the inner
boundary of the forbidden regions, separated from each other in
the left panel and connected by a neck in the right panel.
Radiation has $\omega > 0$ in region ${\cal O}_\i$, divided into
regions ${\cal O}_{\i A}$ and ${\cal O}_{\i B}$ in the right
panel, and $\omega < 0$ in region ${\cal O}_\ii$.}
\label{fig:allIn}}}
\end{figure}

\vskip 2mm Allowed regions for radiation with fixed $\omega$ are
shown in Fig.~\ref{fig:allIn}. As expected, forbidden regions in
plasma are larger as compared with those in vacuum. In the right
panel $q$ exceeds $q_\ct$ defined in Appendix A, therefore
forbidden regions are connected by a neck.

\vskip 5mm {\Large \bf 4. Radiation with fixed $\omega_0$: locally
nonrotating observers}

\vskip 5mm Consider a LNRF observer at $r = r_0$, with the
telescope adjusted to the frequency $\omega_0$. Then, we can use
equation (\ref{eq:om0}) to express $\zeta = \omega_\pl/\omega$ in
terms of $\omega_0$. In this way we obtain
\begin{eqnarray}
\zeta = \pm q \Gamma_0 (1 - \Omega_0 b) \sqrt{\cal R},
\label{eq:zeta}
\end{eqnarray}
where $q$ is the ratio of the plasma frequency on observer's
orbit to the absolute value of the telescope frequency, $q =
\omega_{\pl,0}/|\omega_0|$, $\cal R$ is the electron density ratio
for $r_\r = r_0$, ${\cal R} = N_\e/N_{\e0} = (r_0^2 + r_\c^2)/(r^2
+ r_\c^2)$, and the sign in front of $q$ equals the relative sign
of $\omega$ and $\omega_0$. (For the possibility of $\omega_0 <
0$, see below.) In what follows we assume that $q$ is independent
of $r_0$ so that observers set on different orbits have different
telescope frequencies, the less the larger the radius of the
orbit; in such a way they are able to see the effect of plasma on
the propagation of radiation even at $r \gg 1$, where plasma is
thin. If {\it plasma} is locally nonrotating, we have
$\omega_{\m0} = \omega_0$ and the condition of real-valuedness of
$n$ implies that no ray can reach the observer unless $q < 1$.
However, since $n$ is real in the regions where $F
\ge 0$ regardless of how the plasma moves, the
constraint on $q$ has to be satisfied in general case as well
(as seen also from equation (\ref{eq:Bpm}), where
there appears $1 - q^2$ under the square root).

\vskip 2mm If we insert into $F = \mathring{F} - {\cal D}r^2
\zeta^2$ from (\ref{eq:zeta}), we see that $F$ is quadratic in $b$
just as in the case with fixed $\omega$, when $\zeta = const$;
however, all three coefficients in expression $F = k + 2lb +
mb^2$ now become shifted, not just $k$ (equation (\ref{eq:klm})).
This manifests itself in the behavior of functions $b_\pm$ solving
equation $F = 0$. In particular, the shift in $m$ implies that the
asymptotics of $b_\pm$ at $r \gg 1$ is modified to $b_\pm \sim \pm
Cr$, where $C < 1$, and that the radius $r_\a$ of the asymptote of
function $b_-$ shifts from $r_\a = 2$ to $r_\a < 2$. In Appendix B
we show that $C$ increases from 0 to 1 and $r_\a$ increases from
$r_h$ to 2, as the radius of observer's orbit $r_0$ increases from
$r_h$ to $\infty$. We also discuss there the case when the
function diverging at $r_\a$ is $b_+$ rather than $b_-$.

\vskip 2mm Forbidden regions in the $(r, b)$-plane may be
connected by a neck just as in the case with fixed $\omega$. The
neck is formed in an interval of $r$ where $\Delta = l^2 - km < 0$
(the discriminant of equation $F = 0$ is negative). As shown in
Appendix C, the neck is there for $r_0 \sim r_\h$ as well as for
$r_0 \gg 1$, but it disappears in some interval $\rho_{0A} < r_0 <
\rho_{0B}$. The band connecting forbidden regions
shrinks to a point at some radius $r = \rho_A$, as $r_0$
approaches the value $\rho_{0A}$ from the left, and at some radius
$r = \rho_B > \rho_A$, as $r_0$ approaches the value $\rho_{0B}$
from the right. The values of the radii $(\rho_{0A}, \rho_A)$ and
$(\rho_{0B}, \rho_B)$ are discussed in Appendix C.

\vskip 2mm With $\omega_0$ fixed, we can either restrict to the
rays that actually reach the observer, or to consider also ``ghost
rays'' that bounce on their way back to the observer, but have
such frequency $\omega$ and impact parameter $b$ that their
frequency measured by the observer, provided they reach them,
would have just the desired value. In the latter case we obtain
allowed regions ${\cal O}_\i$ and ${\cal O}_\ii$ similar in form
to those in the problem with fixed $\omega$. The frequency
$\omega_0$, however, must be taken with minus sign for some
``ghost rays'', depending on the value of $b$ when compared to the
limit impact parameter $b_\l = \Omega_0^{-1}$. By (\ref{eq:om0}),
$\omega_0 \propto \omega (1 - b/b_\l)$, and since $\omega > 0$ in
${\cal O}_\i$ and $\omega < 0$ in ${\cal O}_\ii$, we need to put
$\omega_0 < 0$ in the upper part of ${\cal O}_\i$, where $b >
b_\l$, as well as in the lower part of ${\cal O}_\ii$, where $b <
b_\l$.

\begin{figure}[ht]
\centerline{\includegraphics[height=0.32\textheight]{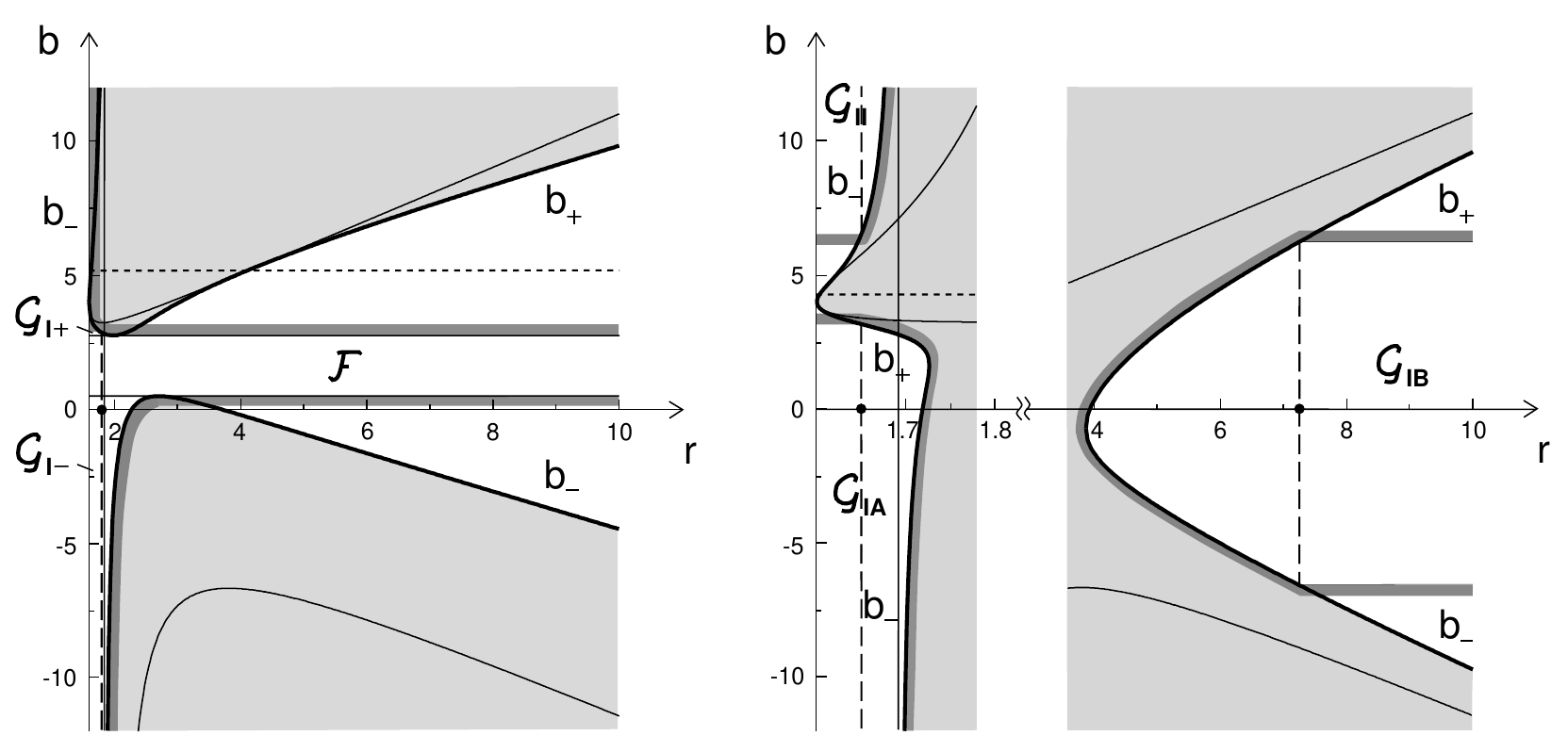}}
\centerline{\parbox {16.2cm}{\caption{\small Allowed regions for
rays in plasma with fixed frequency $\omega_0$ seen by a LNRF
observer. In addition to $a = 0.8$ and $r_\c = 1$ we have $q =
2/3$ and $r_0 =$ 1.8, 1.65 and 7.25 in the left panel, left part
of the right panel and right part of the right panel. Shaded areas
are forbidden regions in an extended theory with ``ghost rays'',
thin lines inside them indicate where the forbidden regions lie in
vacuum, solid vertical line is the asymptote of the inner boundary
of forbidden regions and dotted horizontal line represents $b_\l$,
the value of $b$ dividing regions with ``ghost rays'' in the upper
part of the diagrams into two parts with different sign of
$\omega_0$: in order that $\omega$ has the right sign, $\omega_0$
needs to be $< 0$ above $b_\l$ for $r > 2$ and under $b_\l$ for $r
< 2$ and $b > b_\h$. Position of the observer is marked by a
bullet on the $r$ axis, dashed lines depict rays at the moment of
arriving at the observer and areas within the thick lines, regions
$\cal F$, ${\cal G}_{\i+}$ and ${\cal G}_{\i-}$ in the left panel
and ${\cal G}_{\i A}$, ${\cal G}_\ii$ and ${\cal G}_{\i B}$ in the
right panel, are allowed regions for rays reaching the observer.
Region ${\cal G}_\ii$ in the left panel and line $b_\l$ in the
right part of the right panel did not make it into the diagram
since they lie too high, the former at $b> 25.7$ and the latter at
$b = 276.5$.}\label{fig:allNR}}}
\end{figure}

\vskip 2mm Allowed regions for radiation with fixed $\omega_0$ are
shown in Fig.~\ref{fig:allNR}.
The areas bounded by thick lines are allowed regions for rays that
eventually reach the observer, either directly or after bouncing
back. The remaining free areas outside the shaded corners, or
shaded strip in the case with a neck, are allowed regions for
``ghost rays'' that never arrive at the observer.

\vskip 2mm In the left panel, $r_0$ is in the interval
$(\rho_{0A}, \rho_{0B})$, therefore the corners are separated from
each other and rays can propagate freely from infinity to the
horizon and back within free band $\cal F$. Outside $\cal F$ there
are allowed regions bounded from one side: if $r_0$ is less than
the radius at which $b_+$ has minimum, as in the figure, rays can
reach the observer from the regions ${\cal G}_{\i+}$, ${\cal
G}_{\i-}$ adjacent to $\cal F$ and (not seen in the figure) region
${\cal G}_\ii$ separated from $\cal F$, all bounded from outside.
In the right panel, $r_0$ lies outside the interval $(\rho_{0A},
\rho_{0B})$, hence the corners merge and rays accessible to the
observer are restricted to regions ${\cal G}_{\i A}$ and ${\cal
G}_\ii$ bounded from outside for $r_0 < \rho_{0A}$ (left part of
the panel) and region ${\cal G}_{\i B}$ bounded from inside for
$r_0 > \rho_{0B}$ (right part of the panel). In the case with $r_0
< \rho_{0A}$, we have $q < q_\re$, with $q_\re$ defined in
Appendix B; hence, the asymptote at $r = r_\a$ is approached by
function $b_+$.

\vskip 2mm Rays accessible to  the LNRF observers, who are
orbiting the black hole at different radii $r_0$, occupy regions
$\tilde {\cal O}_\i$ and $\tilde {\cal O}_\ii$ in the $(r_0,
b)$-plane, similar in form to regions ${\cal O}_\i$ and ${\cal
O}_\ii$ in the problem with fixed $\omega$ and $q < q_\ct$ (left
panel of Fig.~\ref{fig:allIn}), with the boundaries $B_\pm$ given
by equation $F(r = r_0) = 0$. As shown in Appendix D, functions
$B_\pm$ are given by the second expression in (\ref{eq:bpm0}),
with $r$ replaced by $r_0$ and with the function under the square
root rescaled as ${\cal D}_0 \to {\cal D}_0/(1 - q^2)$ (equation
(\ref{eq:Bpm})).

\vskip 2mm In the allowed regions in $(r_0, b)$-plane there are
{\it two} rays with the telescope frequency passing through each
point, one directed towards the black hole (with $\dot r_{\ph, 0}
< 0$) and one directed away from it (with $\dot r_{\ph, 0} > 0$).
Let us single out two distinct parts of the allowed regions by the
behavior of rays extended backwards in time, region ${\cal O}_\h$
in which at least one of the rays starts at the horizon and region
${\cal O}_\in$ in which at least one of the rays starts at
infinity (in the sense of limit) -- see Fig.~4. Denote,
furthermore, by ${\cal O}_{\h + \in}$ the intersection of ${\cal
O}_\h$ and ${\cal O}_\in$ and by ${\cal O}_\h'$ and ${\cal
O}_\in'$ the parts of ${\cal O}_\h$ and ${\cal O}_\in$ outside of
${\cal O}_{\h + \in}$. Region ${\cal O}_{\h + \in}$ consists of
rays from the free bands in the $(r, b)$-plane, like band $\cal F$
in the left panel of Fig.~\ref{fig:allNR}, while regions ${\cal
O}_\h'$ and ${\cal O}_\in'$ are composed of rays from allowed
regions bounded from one side, like regions ${\cal G}_{\i+}$ and
${\cal G}_{\i-}$ in the left panel of Fig.~\ref{fig:allNR} and
regions ${\cal G}_{\i A}$, ${\cal G}_\ii$ and ${\cal G}_{\i B}$ in
the right panel of Fig.~\ref{fig:allNR}. If we denote by ${\cal
B}_+$ the upper limit of $\cal F$ (minimum of $b_+$) and by ${\cal
B}_-$ the lower limit of $\cal F$ (maximum of $b_-$ for $r >
r_\a$), we can define ${\cal O}_{\h + \in}$ as a region in the
$(r_0, b)$-plane bounded from above by the line ${\cal B}_+$ and
from below by the line ${\cal B}_-$. Note that the boundaries
actually do not belong to ${\cal O}_{\h + \in}$, since for rays
with $b = {\cal B}_\pm$ the access to the observer is free from
one side only; when arriving from the other side, the rays do not
get past the peak of the forbidden corner standing in their way.
Since the radii $R_\pm$, where $b_+$ has minimum and $b_-$ has
maximum, depend continuously on the radius $r_0$, and since $R
_\pm > r_0 $ at $r_0 = \rho_{0A}$ and $R _\pm < r_0 $ at $r_0 =
\rho_{0B}$, $r_0$ eventually passes through $R_+$ as well as $R_-$
as it rises from $\rho_{0A}$ to $\rho_{0B}$. As a result, lines
${\cal B}_\pm$ touch lines $B_\pm$ in some points $P_\pm$,
dividing the boundary of region ${\cal O}_{\h + \in}$ into two
segments where regions ${\cal O}_\h'$ and ${\cal O}_\in'$ are
attached to region ${\cal O}_{\h + \in}$, ${\cal O}_\h'$ from the
left and ${\cal O}_\in'$ from the right. The position of points
$P_\pm$ as well as the behavior of functions $B_\pm$, ${\cal
B}_\pm$ in their vicinity are discussed in Appendix D.

\vskip 2mm Allowed regions for rays seen by LNRF observers at
various distances from the black hole are shown in
\begin{figure}[ht]
\centerline{\includegraphics[height=0.32\textheight]{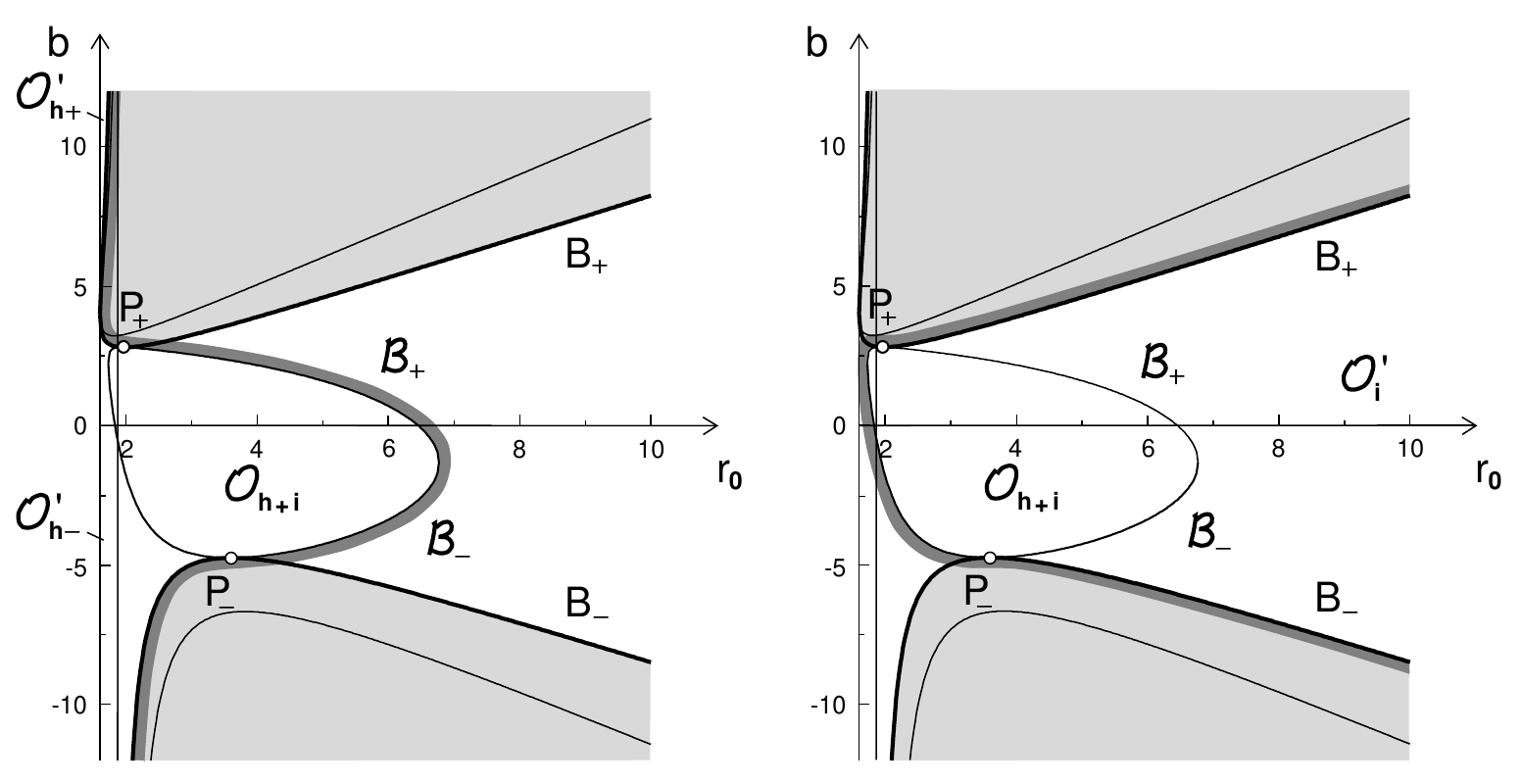}}
\centerline{\parbox {16.2cm}{\caption{\small Allowed regions in
the $(r_0, b)$ plane for radiation seen by LNRF observers.
Parameters $a$, $r_\c$ and $q$ are the same as in
Fig.~\ref{fig:allNR}. Shaded areas are forbidden for rays seen by
the observer at the given radius $r_0$, thin lines inside them are
boundaries of the forbidden regions in vacuum (they look the same
in the $(r_0, b)$ plane as in the $(r, b)$ plane, since the form
of rays in vacuum does not depend on $\omega$), solid vertical
line is the asymptote of the inner boundary of forbidden regions,
region ${\cal O}_{\h + \in}$ in the central part of the diagrams
consists of rays that can reach observers from the horizon as well
as from infinity, regions ${\cal O}_{\h\pm}'$ in the left panel
and region ${\cal O}_\in'$ in the right panel are composed of rays
that can reach observers from one side only, either from the
horizon (regions ${\cal O}_{\h\pm}'$) or and from infinity (region
${\cal O}_\in'$), and points $P_\pm$ divide the boundary of ${\cal
O}_{\h + \in}$ into segments adjacent to these regions. Rays that
can come to the local observers from the horizon form the region
${\cal O}_\h$ to the left of the thick line in the left panel, and
rays that can arrive at the local observers from infinity comprise
the region ${\cal O}_\in$ to the right of the thick line in the
right panel. \label{fig:allNRobs}}}}
\end{figure}
Fig.~\ref{fig:allNRobs}. Region ${\cal O}_\h$, consisting of
${\cal O}_{\h + \in}$ in the center of $\tilde {\cal O}_\i$,
region ${\cal O}_{\h-}'$ adjacent to it and region ${\cal
O}_{\h+}' = \tilde {\cal O}_\ii$ disconnected from both of them,
is shown in the left panel. Region ${\cal O}_\in$, composed of
regions ${\cal O}_{\h + \in}$ and ${\cal O}_\in'$, is displayed in
the right panel. The Kerr parameter is less than critical, $a <
a_\ct$, where $a_\ct$ is defined in Appendix D, therefore the
point $P_+$ is shifted to the left of the static limit. The
boundary of region ${\cal O}_{\h + \in}$ is {\it not} horizontal
at points $P_\pm$ as it may appear from the figure; instead, as
discussed in Appendix D, it is slightly inclined downwards at
$P_+$ and upwards at $P_-$.

\vskip 5mm {\Large \bf 5. Radiation with fixed $\omega_0$: freely
falling observers}

\vskip 5mm As seen from equation (\ref{eq:om0f}), the ratio $\zeta
= \omega_\pl/\omega$ for an observer falling freely from rest at
infinity can be written as
\begin{eqnarray}
\zeta = q \hat \Gamma_0 (1 - \xi_0 \eta_0 - \Omega_0 b) \sqrt{\cal
R}. \label{eq:zetaf}
\end{eqnarray}
The expression is applicable only to rays that can reach the
observer, otherwise the parameter $\eta_0 = \pm \sqrt{F_0}u_0$ is
purely imaginary. Thus, the frequency $\omega_0$ is necessarily
positive; unlike in the case of locally nonrotating observer,
there exist no ,,ghost rays`` with $\omega_0 = - |\omega_0|$. As a
result, parameter $q$ can be defined just as $q =
\omega_{\pl,0}/\omega_0$ and no $\pm$ is needed in equation
(\ref{eq:zetaf}).

\vskip 2mm Function ${\cal F} = Fu^2$ satisfies the identity
${\cal F}_0 = \eta_0^2$, which can be written as a quadratic
equation for $\eta_0$: $K_0 + 2L_0\eta_0 + M_0\eta_0^2 = 0$. We
can define allowed regions $\tilde {\cal O}_\i$ and $\tilde {\cal
O}_\ii$ in the $(r_0, b)$-plane by the condition ${\cal F}_0 \ge
0$ just as for the LNRF observers, however, now this inequality is
not saturated at the {\it boundaries} of $\tilde {\cal O}_\i$ and
$\tilde {\cal O}_\ii$. Therefore, if we want to determine the
shape of these regions, we need to start from the requirement that
equation ${\cal F}_0 = \eta_0^2$ has real solution, i.e., from the
inequality $\delta_0 = L_0^2 - K_0M_0 \ge 0$. As shown in Appendix
E, this leads to the boundaries $\tilde B_\pm$ which are given by
the same expression as the boundaries $B_\pm$ in the problem with
LNRF observers, just with $q$ replaced by $\tilde q_\nr =
q\Gamma_0/\big[1 + q^2 (\Gamma_0^2 - 1)\big]{}^{1/2}$. If we
choose $q$ as a function of $r_0$ so that $\tilde q_\nr = const$,
regions $\tilde {\cal O}_\i$ and $\tilde {\cal O}_\ii$ will have
the same shape for freely falling observers as for locally
nonrotating ones.

\vskip 2mm We have again, as in the problem with LNRF observers,
two rays at each point of regions $\tilde {\cal O}_\i$ and $\tilde
{\cal O}_\ii$; however, the rays now have rescaled radial
velocities $\eta_0^{(\pm)} = (-L_0 \pm \sqrt{\delta_0})/M_0$ which
do not differ just by sign. They can even have the {\it same}
sign: $\eta_0^{(-)}$ is negative in the whole region $\tilde {\cal
O}_\i$, including its boundaries, but $\eta_0^{(+)}$ is positive
in a smaller region, passing through zero at lines $B_\pm$ where
$K_0 = 0$. In region $\tilde {\cal O}_\ii$, on the other hand,
$\eta_0^{(+)}$ is positive everywhere and $\eta_0^{(-)}$ is
negative outside the line $B_-$. However, $\dot r_{\ph,0}$ has
sign different from $\eta_0$ in region $\tilde {\cal O}_\ii$,
since it equals $\omega \eta_0$ and $\omega < 0$ there. Let us
call rays with velocities $\eta_0^{(+)}$ in region $\tilde {\cal
O}_\i$ and $\eta_0^{(-)}$ in region $\tilde {\cal O}_\ii$ the
``rays of class I'', and rays with the other velocities the ``rays
of class II''. As suggested by the discussion of the signs of
$\eta_0^{(\pm)}$, the rays of class II, as well the rays of class
I, lying between the lines $\tilde B_\pm$ and $B_\pm$, are
directed towards the horizon, while the remaining rays of class I
are directed away from it.

\vskip 2mm Should we define regions ${\cal O}_\h$ and ${\cal
O}_\in$ as before, we would have diagrams containing the rays of
both classes I and II, with region ${\cal O}_{\h + \in}$ divided
into three zones, each with rays of different behavior. We will
consider the two classes separately, extending the definition of
regions ${\cal O}_\h$ and ${\cal O}_\in$ by using rays prolonged
in both directions in time: the ray will be supposed to belong to
region ${\cal O}_\h$ if it has, when maximally extended, either
starting or ending point on the horizon, and to region ${\cal
O}_\in$, respectively, if it has either starting or ending point
at infinity.

\begin{figure}[ht]
\centerline{\includegraphics[height=0.32\textheight]{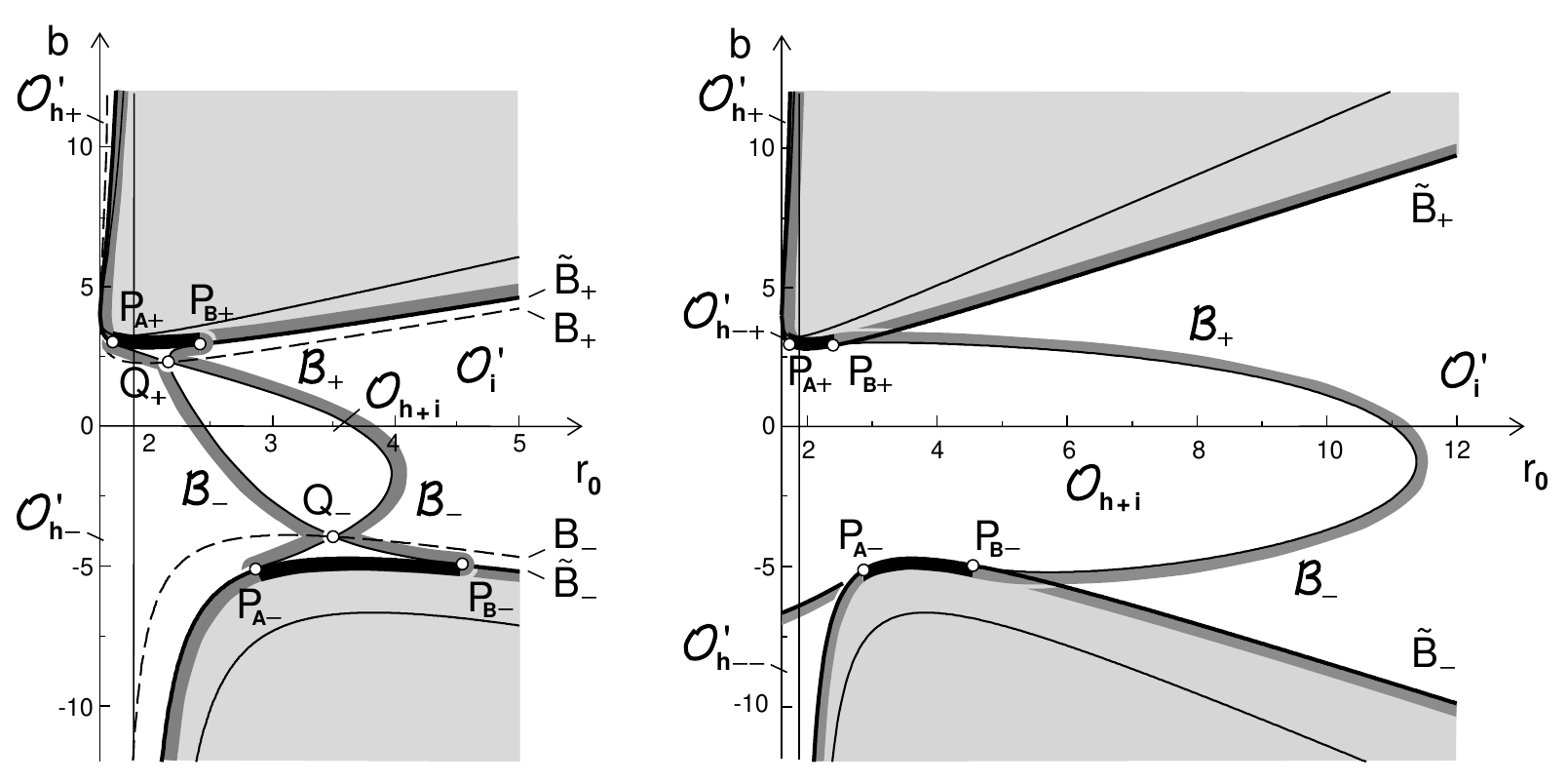}}
\centerline{\parbox {16.2cm}{\caption{\small Allowed regions in
the $(r_0, b)$ plane for radiation seen by freely falling
observers from rest at infinity. Parameter $a$ is the same as in
Fig.~\ref{fig:allV}--\ref{fig:allNRobs}, parameter $r_\c$ is the
same as in Fig.~\ref{fig:allIn}--\ref{fig:allNRobs} and parameter
$\tilde q_\nr$ coincides with $q$ used in Fig.~\ref{fig:allNRobs}.
Allowed regions for rays of class I (with radial velocity
$\eta_0^{(+)}$) and of class II (with radial velocity
$\eta_0^{(-)}$) are shown in the left and right panel,
respectively. The meaning of shaded areas, thin lines inside them
and solid vertical line is the same as in Fig.~\ref{fig:allNRobs},
a new feature are only lines $B_\pm$ (dashed lines in the left
panel), where velocity $\eta_0^{(+)}$ changes its sign. Thick
lines mark, as in Fig.~\ref{fig:allNRobs}, regions ${\cal O}_\h$
and ${\cal O}_\in$; both are now drawn in the same diagram,
boundary of ${\cal O}_\h$ as continuous (``put on the top'') and
boundary of ${\cal O}_\in$ as intermittent (``lying under it'').
For rays of class I, region ${\cal O}_\h$ contains, in addition to
region ${\cal O}_{\h + \in}$, regions ${\cal O}_{\h\pm}'$ to the
left of ${\cal O}_{\h + \in}$, and region ${\cal O}_\in$ contains
region ${\cal O}_\in'$ to the right of ${\cal O}_{\h + \in}$, just
as in Fig.~\ref{fig:allNRobs}; for rays of class II, region ${\cal
O}_{\h + \in}$ extends all the way up to the horizon and region
${\cal O}_{\h-}'$ is divided into regions ${\cal O}_{\h-+}'$ and
${\cal O}_{\h--}'$, adjacent to ${\cal O}_{\h + \in}$ from above
and from below, respectively. Points $P_\pm$, where region ${\cal
O}_{\h + \in}$ has touched forbidden regions in the problem with
LNRF observers, are stretched into finite segments, the same for
rays of both classes (arcs $P_{A+}P_{B+}$ and $P_{A-}P_{B-}$
marked by thick black line). For rays of class I, broken arcs
$P_{A+}P_{A-}$ and $P_{B+}P_{B-}$, forming together the boundary
of ${\cal O}_{\h + \in}$, merge at the points $Q_\pm$ lying on the
lines $B_\pm$.
 \label{fig:allFobs}}}}
\end{figure}

\vskip 2mm Allowed regions for rays seen by freely falling
observers at various distances from black hole are depicted in
Fig.~\ref{fig:allFobs}. In the left panel we show the behavior of
rays of class I, in the right panel -- of class II. Regions ${\cal
O}_\h$ and ${\cal O}_\in$ are displayed both in the same diagram,
marked by continuous and intermittent thick line, respectively. In
Appendix E we explain how we determined those regions without
having an explicit expression for the boundaries of forbidden
regions in the $(r, b)$-plane. If we return to the original
definition of regions ${\cal O}_\h$ and ${\cal O}_\in$, the whole
region ${\cal O}_{\h + \in}$ in the right panel, as well as the
part of region ${\cal O}_{\h + \in}$ lying between the lines
$\tilde B_\pm$ and $B_\pm$ in the left panel, will pass to region
${\cal O}_\in'$, while the rest of region ${\cal O}_{\h + \in}$ in
the left panel will become a part of region ${\cal O}_\h'$.

\vskip 2mm Since the rays with limit impact parameters $\tilde
B_\pm$ have nonzero rescaled radial velocities $\tilde \eta_0 = -
L_0/M_0$, they do not bounce back as they arrive at the observer;
instead, rays coming from infinity bounce back after they pass by
the observer and rays coming from the horizon bounce back before
that (as both have radial velocity $\tilde {\dot r}_{\ph,0} =
\omega \tilde \eta_0 < 0$). In addition, there are also rays
which, rather than bouncing back, proceed directly from infinity
to the horizon. These rays form arcs $P_{A+}P_{B+}$ and
$P_{A-}P_{B-}$, marked by the black segments of thick lines in
Fig.~\ref{fig:allFobs}. Similarly as the points $P_+$ and $P_-$ in
case of LNRF observers, arcs $P_{A+}P_{B+}$ and $P_{A-}P_{B-}$
divide the boundary of region ${\cal O}_{\h + \in}$ into segments
where it borders on regions ${\cal O}_\h'$ and ${\cal O}_\in'$.

\vskip 2mm For rays of class I, allowed region $\tilde {\cal
O}_\i$ between line $\tilde B_+$ and the lower line $\tilde B_-$
contains a smaller region $\tilde {\cal O}_\i'$ between line $B_+$
and the lower line $B_-$, where rays {\it do} bounce
back when arriving at the observer. (As explained in Appendix E,
in order that the latter region extends to the horizon as in
Fig.~\ref{fig:allFobs}, $\tilde q_\nr$ needs to be $<
1/\sqrt{2}$.) At lines $B_\pm$, region ${\cal O}_{\h + \in}$
shrinks to a point. Thus, it is composed of two triangles and a
central part, attached to the boundaries of region $\tilde {\cal
O}_\i'$ at some points $Q_\pm$, just as in the problem with LNRF
observers region ${\cal O}_{\h + \in}$ touched the boundaries of
region $\tilde {\cal O}_\i$ at points $P_\pm$. For rays of class
II, region ${\cal O}_{\h + \in}$ is considerably wider than for
rays of class I; it also does not stop at a finite distance from
the horizon, but extends all the way up. The shape of allowed
regions for rays of both classes I and II is discussed in more
detail in Appendix E.

\vskip 5mm {\Large \bf 6. Conclusions}

\vskip 5mm Using the Hamiltonian approach for tracing rays in
curved spacetime possibly filled with dispersive and refractive
medium we studied their behavior in the Kerr metric. We restricted
to equatorial plane and focused on comparing the form of the
regions accessible to rays in vacuum case with the case when the
black hole is surrounded by cold plasma. We assumed that plasma
has distribution of an isothermal nonsingular sphere with
inhomogeneity parameter comparable with the size of the black hole
and with the density large enough to affect the propagation of
radiation substantially.

\vskip 2mm In the presence of plasma, forbidden regions in the
radius--impact parameter plane, the $(r, b)$-plane, become larger
than in vacuum. They can even develop a neck preventing rays from
passing from infinity to the horizon and back. To understand these
effects, note that the plasma is acting on a ray {\it repulsively}
if its density decreases with radius. Thus, rays are less bent in
plasma than in vacuum, and since extremal values of $b$ at given
$r$ correspond to rays tangent to the circle with radius $r$,
plasma necessarily makes these values less in absolute value in
case the rays are coming from infinity. (If such a ray is
straightened, it needs to enter gravitational field closer to the
radial line parallel to it, in order to get to the same distance
from the source.) The neck connecting forbidden regions is formed
when the repulsive effect is large enough to cause signals coming
from infinity to bounce back, even if they have impact parameter
$b$ close to zero. As for signals coming from the horizon, it
might seem that since they are dragged away from the black hole by
plasma, no matter how small the repulsive effect is near the
horizon, they necessarily escape from the black hole. However, in
the expression for radial acceleration of a light signal there
appears, in addition to a centrifugal acceleration proportional to
$b^2$, the term (with plus sign) proportional to $\dot r^2$. Since
that term is less in the presence of plasma than in vacuum, the
effect of gravity is enhanced by the presence of plasma, and if it
is dense enough, signals coming from the horizon bounce back.

\vskip 2mm We considered also observers located at various radii
$r_0$, equipped with telescopes with a given frequency $\omega_0$,
and constructed allowed regions for rays with that frequency. LNRF
observers were supposed to have frequency $\omega_0$ proportional
to the local plasma frequency, while for freely falling observers
we added an extra factor rising to infinity as the observer's
Lorentzian factor $\Gamma_0$, as the horizon is approached.
Frequency of light signals with the extremal values of $b$
blue-shifts by a factor of order $\Gamma_0$ as we pass from the
LNRF to the freely falling frame, hence it needs to be finite in
the LNRF, just like the telescope frequency with LNRF observers.
The frequency is blue-shifted despite the fact that the observer
is moving away from the source, because of the transversal Doppler
effect. However, rays coming from infinity can reach the horizon
only if their impact parameter is separated by a finite gap from
the maximum $b$ in the region with positive frequency at infinity
$\omega$; and with such $b$, rays with the velocity directed to
the black hole are redshifted rather than blueshifted by the
transition from the LNRF to the observer's frame. Their redshift
is quite substantial, of order $\Gamma_0^{-1}$, so that their
frequency in the LNRF is of the same order of magnitude as the
Lorentzian factor $\hat \Gamma_0 = \Gamma_0^2$ of a freely falling
observer.

\vskip 2mm In the diagrams with multiple observers we constructed
regions in which at least one of the two rays passing through each
point of the diagram started at the horizon (region ${\cal O}_\h$)
or at infinity (region ${\cal O}_\in$). The setting with freely
falling observers is more general than that with LNRF observers,
since the rays seen by them are independent of each other. For
LNRF observers, the rays are related by a simultaneous change of
sign of coordinates $t$ and $\phi$; thus, they are mirror images
of each other with respect to the radial line drawn through the
point of observation, while their velocities are mirror images of
each other with respect to the {\it tangential} line drawn through
that point. As a result, the starting point of each ray, obtained
by extending it maximally backwards in time, coincides with the
ending point of the other ray, obtained by extending it maximally
{\it forwards} in time, provided it passed by the observer without
being detected. For freely falling observers the two rays are not
mirror images of each other, since the radial motion of the
observer breaks the symmetry with respect to the tangential line.
Thus, we have two classes of rays that are {\it not} related by
symmetry, one containing a subclass of rays directed away from the
black hole and the other consisting entirely of rays directed
towards the black hole. By extending the rays in both directions
in time we obtain a separate diagram for each class. The diagrams
look quite different from the one constructed with LNRF observers:
the intersection of regions ${\cal O}_\h$ and ${\cal O}_\in$ is a
narrow strip with ragged boundary, located close to the horizon,
for rays of the first class, and a wide band with smooth boundary,
reaching far from the horizon on one side and adjacent to the
horizon on the other side, for rays of the second class.

\vskip 2mm Regions ${\cal O}_\h$ and ${\cal O}_\in$ arise in the
diagrams with multiple observers because of the presence of
plasma; they reflect the fact that if the plasma is sufficiently
dense, there may appear necks in the diagrams with one observer.
No such regions exist for rays in vacuum, whose form does not
depend on frequency. Thus, the shape of regions ${\cal O}_\h$ and
${\cal O}_\in$ may be regarded as a visualization of the interplay
between strong gravitational field and plasma as an example of a
refractive and dispersive medium, as they act on radiation
propagating through them.

\vskip 5mm

{\Large \bf Declarations}

\vskip 2mm {\it Acknowledgements.} V. B. was supported by the
grant VEGA 1/0719/23, B. B. acknowledges the support of the
Charles University Grant Agency under the Contract No. 317421. B.
B. and J. B. were supported by the Czech Grant Agency under the
Contract No. 21/112685.

\vskip 2mm {\it Competing interests.} The authors have no relevant
financial or non-financial interests to disclose.

\vskip 5mm

{\Large \bf A. Formation of neck in the case with fixed $\omega$}

\vskip 5mm For radiation with fixed frequency $\omega$, let us
find the condition for the formation of a neck between forbidden
regions. Functions $b_\pm$ in the case with fixed $\omega$ are
\begin{equation} \tag{A.1}
b_\pm = \frac 1f (-2au \pm \sqrt{\Delta}), \quad \Delta = (1 -
q^2f{\cal R}){\cal D}r^2.
 \label{eq:bpm}
\end{equation}
As mentioned in Section 3, this is just the first expression in
(\ref{eq:bpm0}) with $\cal D$ multiplied by $1 - f\zeta^2$ (since
$\zeta^2 = q^2\cal R$). Function $\hat{\cal R} = f{\cal R}$
appearing in the definition of $\Delta$ can be written as
$$\hat{\cal R} \propto (1 - 2u)u^2/(1 + r_\c^2 u^2).$$
The function is positive for $0 < u < 1/2$, it vanishes at both $u
= 0$ and $u = 1/2$, and its derivative, $d\hat{\cal R}/du \propto
1 - 3u - r_\c^2 u^3$, decreases monotonically with $u$, crossing
zero at some $u_C < 1/2$. Thus, $\hat{\cal R}$ has the maximum at
$r_C = u_C^{-1} > 2$. Denote $q_\ct = \hat{\cal R}_C^{-1/2}$. For
$q = q_\ct$, the curves $b_+$ and $b_-$ merge at $r_C$; in fact,
they develop spikes directed against each other, whose tips touch
at $r_C$. The reason is that in the vicinity of $r = r_C$ the
function $\Delta$ is proportional to $\delta^2$, where $\delta = r
- r_C$, so that functions $b_\pm$ assume the form $b_\pm = A +
B\delta \pm C|\delta|$, and as a result, have spike-like minimum
and maximum at $\delta = 0$. For $q > q_\ct$, $\Delta$ becomes
negative in some interval $(r_A, r_B)$ containing $r_C$, the
larger the greater the value of $q$; therefore, the two regions
inaccessible for rays become connected by a neck that extends from
$r_A$ to $r_B$. For the parameters used in Fig.~\ref{fig:allIn} we
have $r_C = 3.104$, which yields $\hat{\cal R}_C = 0.1418$ and
$q_\ct = 2.656$. Note that the interval $(r_A, r_B)$ lies entirely
above $r = 2$ for any $q > q_\ct$, since $\Delta = (1 +
|f|\zeta^2){\cal D}r^2 > 0$ for $r < 2$. For $q = 3.5$ (the value
used in the right panel of Fig.~\ref{fig:allIn}), the neck extends
from $r_A = 2.268$ to $r_B = 5.727$.

\vskip 5mm {\Large \bf B. Allowed regions far from the horizon and
close to it in the case with fixed $\omega_0$}

\vskip 5mm For radiation with fixed frequency $\omega_0$ seen by a
LNRF observer, function $F$ determining allowed regions has the
form $F  = k + 2lb + mb^2$, where coefficients $k$, $l$, $m$ can
be expressed in terms of function $Q = q\Gamma_0\sqrt{\cal DR}r$
as
\begin{equation} \tag{B.1}
k = {\cal A}r^2 - Q^2, \quad l = -2au + Q^2 \Omega_0, \quad m = -
f - Q^2 \Omega_0^2.
 \label{eq:klm}
\end{equation}
Solutions to equation $F = 0$ are $b_\pm =$ $-(l \pm
\sqrt{\Delta}$ \hskip -1mm$)/m$, where $\Delta = l^2 - km$.

\vskip 2mm For $r \gg 1$, $Q$ approaches the value $Q_\infty =
q\Gamma_0r_{0+}$, where $r_{+} = (r^2 +
r_\c^2)^{1/2}$; hence, $F$ can be written as $F
\doteq r^2 + 2Q_\infty^2\Omega_0b - (1 + Q_\infty^2\Omega_0^2)b^2$
and the solutions to the equation $F = 0$ in the leading order in
$r$ are $b_\pm = \pm Cr$, where $C = (1 + Q_\infty^2
\Omega_0^2)^{-1/2}$. Since $C < 1$, the wedge-like band between
the forbidden regions narrows down if we fix $\omega_0$ instead of
$\omega$.

\vskip 2mm In order to determine the behavior of constant $C$ as a
function of $r_0$, let us express constant $\hat q =
Q_\infty\Omega_0$ appearing in the definition of $C$ as
$$\hat q \propto r_{0+}u_0^3/\sqrt{{\cal A}_0
{\cal D}_0} = r_{0+}/(\sqrt{r_0 A_0 D_0}),$$ where $A = {\cal
A}r^3 = r^3 + a^2r + 2a^2$ and $D = {\cal D}r^2 = r^2 -2r + a^2$.
The expression for $\hat q$ can be written as a product of
functions $r_+/r$, $\sqrt{r/A}$ and $1/\sqrt{D}$, evaluated at $r
= r_0$; and since function $r_+/r$ obviously decreases with $r$
for all $r > 0$, while functions $A/r$ and $D$ can be shown to
increase with $r$ for $r > r_\h$, constant $\hat q$ decreases and
constant $C = (1 + \hat q^2)^{-1/2}$ increases as $r_0$ goes from
$r_\h$ to $\infty$. The former constant decreases from $\infty$ to
0 for $r_0$ increasing from $r_\h$ to $\infty$, hence, the latter
constant increases from 0 to 1. For the parameters used in
Fig.~\ref{fig:allNR} we have $C =$ 0.681 (left panel) and 0.9997
(right part of the right panel).

\vskip 2mm Consider now the radius $r_\a$ at which either $b_-$ or
$b_+$ has an asymptote. The value of $r_\a$ is given by the
equation $m_\a = -f_\a - Q_\a^2 \Omega_0^2 = 0$, i.e., $f_\a =
-\hat q^2 D_\a/r_{\a+}^2$ (since $Q = Q_\infty \sqrt{D}/r_+$).
Function $f$ is negative only for $r < 2$ so that the asymptote
lies necessarily there; and since it is located at $r = 2$ in case
with fixed $\omega$, the inner borderline of forbidden regions, or
region in case with a neck, shifts towards $r_\h$ as we fix
$\omega_0$ instead of $\omega$.

\vskip 2mm Function $-f$ falls from a positive value to 0 as $r$
rises from $r_\h$ to 2, while function $ D/r_+^2$ can be shown to
rise from 0 to a positive value. Thus, the asymptote is the closer
to the static limit the smaller the value of $\hat q$; and since
$\hat q$ falls from $\infty$ to 0 as $r_0$ rises from $r_\h$ to
$\infty$, the asymptote shifts with increasing $r_0$ from $r_\h$
to 2. For the parameters used in Fig.~\ref{fig:allNR} we have
$r_\a =$ 1.836 (left panel) and 1.691 (the left part of right
panel).

\vskip 2mm For radiation with fixed $\omega$ we had $r_\a = 2$ and
the divergent function was $b_-$, just as in case without plasma.
Now we have $r_\a < 2$ and the divergent function may be $b_+$,
too: it is $b_-$ if $l_\a < 0$ and $b_+$ if $l_\a > 0$. In the
latter case, forbidden regions are necessarily connected with a
neck and the way their boundary approaches the asymptote at $r =
r_\a$ is reversed as compared with the former case: function $b_+$
is falling to $-\infty$ to the left of the asymptote and rising to
$+\infty$ to the right. Let us write $l_\a = -2au_\a -
f_\a/\Omega_0 \propto \lambda_\a - 1$, where $\lambda =
-f/(2a\Omega_0u) = A_0 (2 - r)/(4a^2)$. As $r$ rises from $r_\h$
to 2, $\lambda$ falls linearly from $\lambda_\h = A_0
r_{\h-}/(4a^2)$ to 0, where $r_{\h-}$ is the radius of the inner
horizon,  $r_{\h-} = 1 - \sqrt{1 - a^2}$ (the smaller root of
${\cal D} = 0$). It can be easily proven that $\lambda_\h > 1$, so
that $\lambda > 1$ below some radius $r_1 > r_\h$; explicitly,
$r_1 = r_\h + r_{\h-} (1 - 1/\lambda_\h)$. For the parameters used
in the left part of the right panel of Fig.~\ref{fig:allNR} we
have $r_1 = 1.625 < r_\a = 1.691$; thus, the function diverging at
$r_\a$ is $b_-$. Note that in order that the diverging function is
$b_+$, the parameter $q$ needs to satisfy $q
> q_\re = (\Gamma_0\Omega_0)^{-1} \sqrt{-f_1/(D_1{\cal R}_1)}$
(``rev'' stands for ``reversed''). In our case $q_\re
= 1.421$, so that $q < q_\re$.

\vskip 5mm {\Large \bf C. Formation of neck in case with fixed
$\omega_0$}

\vskip 5mm In the case radiation has fixed frequency $\omega_0$
with respect to a LNRF observer, the discriminant of equation $F =
0$ is
\begin{equation} \tag{C.1}
\Delta = (1 - q^2 \Gamma_0^2 f_+{\cal R}){\cal D}r^2, \quad f_+ =
f + 4ar^{-1}\Omega_0 - {\cal A}r^2 \Omega_0^2.
 \label{eq:Del}
\end{equation}
If forbidden regions in the $(r, b)$-plane are connected by a
neck, the radii $r_A$ and $r_B$, where the inner edge of the neck
is farthest from the horizon and the outer edge is closest to it,
are given by the equation $X = 0$, where $ X = 1 - q^2 \Gamma_0^2
f_+{\cal R}$ is the extra factor by which $\Delta$ differs from
$\mathring{\Delta} = {\cal D}r^2$ (see equation~(\ref{eq:Del})).
While in case with fixed $\omega$ the neck forms only for large
enough $q$, now it can exist for arbitrary $q$, provided $r_0$
lies outside a certain interval $(\rho_{0A}, \rho_{0B})$. To see
this, consider asymptotics of $X$ for $r_0 \sim r_\h$ and $r_0 \gg
1$.

\vskip 2mm Close to the horizon, at $\epsilon = r - r_\h \ll 1$,
we have ${\cal D} \doteq 2\sqrt{1 - a^2} u_\h^2 \epsilon \equiv
u_\h^2 \hat \epsilon$, and the value of function $\cal A$ at the
horizon is ${\cal A}_\h = 4u_\h^2$. Thus, if the observer is
shifted w.r.t. the horizon by $\epsilon_0 \ll 1$, their
$\Gamma$-factor is $\Gamma_0 \doteq ({\cal A}_\h/{\cal D}_0)^{1/2}
\doteq 2\hat \epsilon_0^{-1/2}$; furthermore, for the angular
velocity of the observer we have $\Omega_0 \doteq \Omega_\h  =
\frac 12 au_\h$. As seen from these expressions, function $X$
reduces close to the horizon to $X \doteq 1 - 4q^2 \hat
\epsilon_0^{-1}f_+$, where $rf_+ = r - 2 + 4a\Omega_0 (1 - \frac
14 a^{-1}A\Omega_0) \doteq r - 2 + 2a^2u_\h (1 - \frac 18 u_hA)
\doteq [1 - \frac 14 a^2 (3 + a^2u_\h^2)]\epsilon \equiv
C\epsilon$. (We have used the fact that if $r_0$ as well as $r$
are $\sim r_\h$, then ${\cal R} \doteq 1$.) As a result, for
$\epsilon_0 \ll 1$ there exists a neck in the $(r, b)$-plane with
the inner edge bent away from the horizon up to $\epsilon_A =
\frac 14 q^{-2} (r_\h/C) \hat \epsilon_0$. If we denote $\alpha =
\sqrt{1 - a^2}$, we have $\hat \epsilon_0 = 2\alpha \epsilon_0$
and $4r_\h^2 C = (4 - 3a^2)r_\h^2 - a^4 = 2\alpha r_\h^3$ (as seen
by writing $r_\h$ as $1 + \alpha$); hence, $\epsilon_A =
q^{-2}\epsilon_0$. Note that from the expansion of $F$ up to the
first order in $\epsilon$ it follows that the asymptote of $b_-$
is shifted w.r.t. the horizon by the same value, $\epsilon_a =
q^{-2}\epsilon_0$. Thus, if we want to check that $\epsilon_A >
\epsilon_a$ (the point where $b_-$ merges with $b_+$ lies beyond
the asymptote of either $b_-$ or $b_+$), we need to proceed to the
next-to-leading order in $\epsilon_0$.

\vskip 2mm  In the opposite limit, when $r_0 \gg 1$, we have
$\Gamma_0 \doteq 1$ and $\Omega_0 = O(u_0^3)$; if, moreover, $r
\sim r_0$, we have also $f_+ \doteq 1$ and ${\cal R}\doteq
r_0^2u^2$. Thus, the asymptotics of function $X$ far from the
horizon is $X \doteq 1 - q^2r_0^2u^2$; and, consequently, there
exists a neck in the $(r, b)$-plane with the outer edge bent
towards the horizon down to $r_B = qr_0$.

\vskip 2mm The values of $r_0$ for which the neck shrinks to a
point, along with the values of $r$ where the point is located,
are given by equations $X = X'= 0$. The first equation is just
$q^2 \Gamma_0^2 f_+{\cal R} = 1$, while the second equation can be
written as the cubic equation for $u = r^{-1}$,
$$(f_+{\cal R})' \propto (1 - a\Omega_0)^2u(3 + r_\c^2u^2) -
(r_\c^2 - a^2)\Omega_0^2 - 1 = 0.$$ If we express $u$ as function
of $\Omega_0$ by Cardano's formula and solve equation $X = 0$ with
this $u$ inserted into it numerically, we obtain two pairs of
values $(r_0, r)$. For the parameters used in
Fig.~\ref{fig:allNRobs} these values are $(\rho_{0A}, \rho_A) =
(1.726, 2.176)$ and $(\rho_{0B}, \rho_B) = (6.769, 3.080)$. As
seen from our analysis of the limit cases $r_0 \sim r_\h$ and $r_0
\gg 1$, forbidden regions develop a neck for $r_0 < \rho_{0A}$ as
well as for $r_0 > \rho_{0B}$; hence, for $r_0$ from the interval
$(\rho_{0A}, \rho_{0B})$ they are necessarily separated by a free
band. Moreover, the fact that the interval $(\rho_A, \rho_B)$ lies
entirely inside the interval $(\rho_{0A}, \rho_{0B})$ suggests
that the inequalities $\epsilon_A = q^{-2}\epsilon_0 > \epsilon_0$
and $r_B = qr_0 < r_0$, obtained in the limits $r_0 \sim r_\h$ and
$r_0 \gg 1$, stay valid all the way from $r_0 = r_\h$ up to $r_0 =
\rho_{0A}$ and from $r_0 = \infty$ down to $r_0 = \rho_{0B}$. This
is easily verified numerically, because equation $X = 0$ is cubic
in $r$, therefore we can once again use Cardano's formula to
express its solutions, radii $r_A$ and $r_B$, analytically as
functions of $r_0$.

\vskip 5mm {\Large \bf D. Constructing allowed regions in the
$(r_0, b)$-plane}

\vskip 5mm Let us find functions $B_\pm = b_\pm (r_0)$ for a LNRF
observer. Define ${\cal F} = Fu^2 = \mathring{\cal F} - {\cal
D}\zeta^2$, where $\mathring{\cal F} = \mathring{F}u^2 = {\cal A}
- 4au^3 b - fu^2 b^2$ and $\zeta$ is given by equation
(\ref{eq:zeta}). If we rewrite $\mathring{\cal F}$ as
$\mathring{\cal F} = {\cal A}(1 - \Omega b)^2 - {\cal D}u^2
b^2/{\cal A}$, we obtain $\mathring{\cal F}_0 = {\cal
A}_0\kappa_0^2 - {\cal D}_0u_0^2 b^2/{\cal A}_0$, where $\kappa_0
= 1 - \Omega_0 b$. [The expression for $\mathring{\cal F}$ which
we have started from follows immediately from the identity $f{\cal
A} + 4a^2u^4 = {\cal D}$, but it can also be obtained by rewriting
the 2D metric as $ds^2_2 = -{\cal D}dt^2/{\cal A} + {\cal
A}r^2d\tilde\phi^2$, where $d\tilde\phi = d\phi - \Omega dt$, and
by regarding the formula $\mathring{\cal F} = -\omega^{-2}{\cal
D}\tilde g^{AB} \tilde k_A \tilde k_B$ with $\tilde k_A = \omega
(-1 + \Omega b, b)$.] Since $\zeta_0^2 = q^2\Gamma_0^2\kappa_0^2 =
q^2{\cal A}_0\kappa_0^2/{\cal D}_0$, we find ${\cal F}_0 = (1 -
q^2){\cal A}_0\kappa_0^2 - {\cal D}_0u_0^2 b^2/{\cal A}_0 \propto
\mathring{\cal F}_0({\cal D}_0 \to {\cal D}_0/(1 - q^2))$, and by
putting this expression equal to zero we obtain
\begin{equation} \tag{D.1}
B_\pm = \frac {{\cal A}_0r_0^2}{2ar_0^{-1} \pm \sqrt{{\cal D}_0/(1
- q^2)}r_0}.
 \label{eq:Bpm}
\end{equation}

\vskip 2mm Functions $B_\pm$ have the asymptotics $B_\pm =$ $\pm
\sqrt{1 - q^2}r_0$ for $r_0 \gg 1$. Furthermore, the asymptotics
of ${\cal F}_0$ for $b \to \pm \infty$ is ${\cal F}_0 = \big[(1 -
q^2){\cal A}_0\Omega_0^2 - {\cal D}_0u_0^2/{\cal A}_0\big]b^2
\propto \big[1 - q^2 - {\cal D}_0 r_0^4/(4a^2) \big]b^2$; hence,
the radius $r_{0\a}$ at which function $B_-$ has an asymptote is
given by $\sqrt{{\cal D}_{0\a}}r_{0\a}^2/(2a) = \sqrt{1 - q^2}$.
For the parameters used in Fig.~\ref{fig:allNRobs} we have
$r_{0\a} = 1.874$.

\vskip 2mm For any given $r_0$ from the interval $(\rho_{0A},
\rho_{0B})$ we can find the impact parameters ${\cal B}_\pm$ at
the lowest point of the upper forbidden corner and at the highest
point of the lower forbidden corner in the $(r, b)$-plane, and by
repeating this procedure for a sequence of nearby $r_0$'s covering
the interval $(\rho_{0A}, \rho_{0B})$ we can construct the lines
${\cal B}_\pm$ in the $(r_0, b)$-plane. The line ${\cal B}_-$ lies
under the line ${\cal B}_+$ inside the interval $(\rho_{0A},
\rho_{0B})$ and merges with it at the end points of the interval;
therefore, the two lines form a loop in the $(r_0, b)$-plane. In
region ${\cal O}_{\h + \in}$ inside the loop, rays can reach
observers from both sides, from the horizon as well as from
infinity, while at the boundary of ${\cal O}_{\h + \in}$ (the line
$\partial {\cal O}_{\h + \in}$ composed of ${\cal B}_+$ and ${\cal
B}_-$) the rays arrive at the observers from one side only, either
from the horizon or from infinity. Let us discuss the position of
the points $P_\pm$ separating the two parts of the $\partial {\cal
O}_{\h + \in}$ with opposite directions of the incoming light, and
the behavior of the lines $B_\pm$ and ${\cal B}_\pm$ in their
neighborhood.

\vskip 2mm As $r_0$ rises from $\rho_{0A}$ to $\rho_{0B}$, the
radii $R_+$ and $R_-$, at which function $b_+$ has minimum (radius
$R_+$) and function $b_-$ has maximum (radius $R_-$), change from
$\rho_A$ to $\rho_B$: the radius $R_+$ first rises and then falls,
while the radius $R_-$ first falls and then rises. Since $\rho_A >
\rho_{0A}$ and $\rho_B < \rho_{0B}$, both radii necessarily equal
$r_0$ at some point, the radius $R_+$ at $r_0 = R_{0+}$ and the
radius $R_-$ at $r_0 = R_{0-}$. By the definition of points
$P_\pm$, the radii $R_{0\pm}$ are the values of the radius $r_0$
at which these points are located in the $(r_0, b)$-plane.
Obviously, functions $B_+ (r_0)$ and ${\cal B}_+ (r_0)$ coincide
at $r_0 = R_{0+}$, and functions $B_- (r_0)$ and ${\cal B}_-
(r_0)$ coincide at $r_0 = R_{0-}$. The radii $r = R_\pm$ are given
by equation $\partial_r b_\pm(r_0, r) = 0$ or, equivalently, by
equation $\partial_r F(r_0, r, b)\big|_{b = b_\pm(r_0, r)} = 0$.
Thus, for radii $r_0 = R_{0\pm}$ we have equation $\partial_r
F(r_0, r_0, b)\big|_{b = B_\pm(r_0)} = 0$. To put this into a
compact form, introduce new variables $x = fb + 2au$ and $y = 1 -
\Omega_0 b$ and write functions $F$ and $\partial_r F$ as $fF = D
- x^2 - fQ^2y^2$ and $\partial_r (fF) \propto \frac 12 \mu - (x -
\frac 12 a)^2 - \frac 12 \hat \nu y^2$, where $Q^2 = q^2\Gamma_0^2
D\cal R$, $\mu = \frac 12 (fD'r^2 + a^2) = r^3 - 3r^2 + 2r + \frac
12 a^2$ and $\hat\nu = \frac 12 (fQ^2)' = q^2\Gamma_0^2 (r^3 -
2r^2 + a^2 - fDr^3/r_+^2) \equiv q^2\Gamma_0^2 \nu$. By combining
equations $F(r_0, r_0, b) = 0$ (definition of $B_\pm$) and
$\partial_r F(r_0, r_0, b) = 0$, we obtain
\begin{equation} \tag{D.2}
\Phi_\pm \equiv \tfrac 12 \mu_0 - (X_\pm - \tfrac 12 a)^2 - \tfrac
12 \nu_0 (1 - X_\pm^2/D_0) = 0,
 \label{eq:Ph}
\end{equation}
where $X_\pm = x_0(B_\pm) = f_0 B_\pm + 2au_0$. The desired radii
are found by solving equations $\Phi_\pm = 0$ for $r_0$; for the
parameters used in Fig.~\ref{fig:allNRobs} the radii are $R_{0+} =
1.965$ and $R_{0-} = 3.593$.

\vskip 2mm Points $P_\pm$ are not just common points of lines
${\cal B_\pm}$ and $B_\pm$; they are {\it points of the tangency}
of those lines. This is clear from the very fact that there is
just one common point for each pair of lines, but it can be seen
also from the behavior of functions ${\cal B}_\pm (r_0)$ and
$B_\pm (r_0)$ close to $R_{0\pm}$. Consider functions with index
$+$. Since ${\cal B}_+ (r_0)$ is the minimum of function $b_+(r_0,
r)$, there must hold $B_+ > {\cal B}_+$ in the neighborhood of
$R_{0+}$; however, this can be satisfied for $r_0
> R_{0+}$ only if $dB_+/dr_0 \ge d{\cal B_+}/dr_0$ at $R_{0+}$,
and for $r_0 < R_{0+}$ only if $dB_+/dr_0 \le d{\cal B}_+/dr_0$ at
$R_{0+}$. Thus, the derivatives of $B_+$ and ${\cal B}_+$ coincide
at $R_{0+}$.

\vskip 2mm Equation $\Phi_\pm = 0$ has, in addition to solutions
$r_0 = R_{0\pm}$ depending on parameters $a$, $r_\c$ and $q$, also
the solution $r_0 = 2$ valid for all values of parameters. Indeed,
for $r_0 = 2$ we have $\mu_0 = \frac 12 a^2$, $X_\pm
= a$ and $D_0 = a^2$, so that $\Phi_\pm = \frac 14 a^2 - (a -
\frac 12 a)^2 = 0$. However, this does not mean that the radii
$R_\pm$ merge with $r_0$ twice, once at the static limit, provided
it falls into the interval $(\rho_{0A}, \rho_{0B})$, and once
outside it. The reason is that the coefficient of proportionality
in the expression for $\partial_r F$ contains factor $f^{-2}$, so
that the equation to be solved is actually $f_0^{-2} \Phi_\pm =
0$; and since for $r_0 \sim 2$ it holds $f_0 = \epsilon_0/2$ and
$\Phi_\pm = O(\epsilon_0^2)$, where $\epsilon_0 = r_0 - 2$, the
equation is in the generic case {\it not} solved by $r_0 = 2$. If
we introduce constants $k_\pm = (p \mp 1)/(p \pm 1)$, where $p =
1/\sqrt{1 - q^2}$, and denote $l = 2/r_+(2) = 2/(4 +
r_\c^2)^{1/2}$, we can write the coefficients in $\Phi_\pm = K_\pm
\epsilon_0^2$ as
\begin{equation*}
K_\pm = -(1 + \tfrac 12 a^2) \big[2(1 + \tfrac 12 a^2)a^{-2}q^{-2}
- l^2 - \tfrac 12\big]k_\pm + \tfrac 32 +\tfrac 18 a^2.
 \label{eq:Ph1}
\end{equation*}
The coefficients are both $\propto -a^{-2}$ for $a \sim 0$, and as
$a$ rises to 1, $K_-$ stays negative, while $K_+$ crosses 0 at
some $a_\ct < 1$, provided $q$ is not too close to 1. If
parameters $r_\c$ and $q$ are such as in Fig. 4, $r_\c = 1$ and $q
= 2/3$, the critical value of $a$ is $a_\ct = 0.784$. From the
behavior of functions $\Phi_\pm$ at $r_0 = r_\h$ and $r_0 \gg 1$
it follows that they cross 0 at $r_0 > 2$ if they touch the $r_0$
axis from below at $r_0 = 2$, and at $r_0 < 2$ if they touch the
$r_0$ axis from above. Thus, $R_{0-} > 2$ for all $a$'s and
$R_{0+}> 2$ for $a < a_\ct$, but $R_{0+}< 2$ for $a > a_\ct$. The
value of $a$ used in Fig. 4 is a bit greater than $a_\ct$, that is
why point $P_+$ is shifted a bit to the left of $r_0 = 2$.

\vskip 2mm The derivatives of $B_\pm$ and ${\cal B}_\pm$ at $r_0 =
R_{0\pm}$ are proportional to the function
$$(\partial_{r_0} F)_0 = -2Q_0^2 Yy, \quad Y =
[(r_0 - a^2u_0^2)r_0/A_0 - (r_0 - 1)/D_0 + r_0/r_{0+}^2]y +
2a(3r_0^2 + a^2)b/A_0^2,$$ evaluated at $b = B_\pm$. The
derivatives seem to be zero in Fig. 4, but in fact they are only
{\it almost} zero. To see this, compare the radii $R_{0\pm}$ to
the radii $\hat R_{0\pm}$ at which $Y = 0$: for the parameters
used in Fig. 4 it holds $\hat R_{0+} = 1.944$ and $\hat R_{0-} =
3.580$; hence, $\hat R_{0+}$ is close to $R_{0+}$ and $\hat
R_{0-}$ is close to $R_{0-}$. By combining the expression for
$(\partial_{r_0} F)_0$ given above with the formula $\partial_b
F_0 = -2(x_0 - Q_0^2 \Omega_0 y)$, we find that $d{\cal B}/dr_0 =
-0.0794$ at $r_0 = R_{0+}$ and $d{\cal B}/dr_0 = 0.0074$ at $r_0 =
R_{0-}$.

\vskip 5mm {\Large \bf E. Allowed regions for freely falling
observers}

\vskip 5mm We are interested in allowed regions in the $(r_0, b)$
plane for rays seen by freely falling observers. Function ${\cal
F} = Fu^2$ determining these regions has again, as in the problem
with LNRF observers analyzed in Appendix D, the form ${\cal F} =
\mathring{\cal F} - {\cal D}\zeta^2$; however, $\zeta$ is now
given by equation (\ref{eq:zetaf}). Let us write the additional
term appearing in $\cal F$ due to the presence of plasma as ${\cal
D}\zeta^2 = \gamma (\kappa_0 - \xi_0\eta_0)^2$, where $\kappa_0 =
1 - \Omega_0 b$ and $\gamma = q^2\hat \Gamma_0^2 {\cal D}{\cal
R}$. (If we wrote function $\cal F$ for a LNRF observer as
quadratic polynomial in $b$, there would appear, in the
coefficients instead of function $Q^2$, function ${\cal Q}^2 = Q^2
u^2 = q^2\Gamma_0^2 {\cal D}{\cal R}$. Since $\gamma$ is obtained
from this function by replacing $\Gamma_0 \to \hat \Gamma_0$, we
can write it as $\gamma = {\hat {\cal Q}}^2$.) To make formulas
describing radiation seen by the two classes of observers more
similar to each other, define an ``effective LNRF parameter''
$q_\nr = q\hat \Gamma_0/\Gamma_0 = q\Gamma_0$; then, $\gamma =
q_\nr^2 \Gamma_0^2 {\cal DR}$ and $\gamma_0 = q_\nr^2 {\cal A}_0$.

\vskip 2mm A new feature of the theory in the case with freely
falling observers is the identity ${\cal F}_0 = \eta_0^2$. The
difference between the two sides of the equation can be written as
the polynomial $K_0 + 2L_0\eta_0 + M_0\eta_0^2$, where
\begin{equation} \tag{E.1}
K_0 = - \mathring{\cal F}_0 + \gamma_0 \kappa_0^2, \quad L_0 =
-\gamma_0\xi_0\kappa_0, \quad M_0 = 1 + \gamma_0 \xi_0^2.
 \label{eq:KLM}
\end{equation}
The polynomial has two roots $\eta_0^\br = (-L_0 \pm
\sqrt{\delta_0})/M_0$, where $\delta_0 = L_0^2 - K_0 M_0$. Note
that the coefficient in front of $\eta_0^2 $ can be written as
$M_0 = 1 + q_\nr^2 (1 - \Gamma_0^{-2})$.

\vskip 2mm By plugging expressions for $K_0$, $L_0$, $M_0$ into
the definition of $\delta_0$ we find $\delta_0 = M_0
\mathring{\cal F}_0 - \gamma_0 \kappa_0^2$. If we define a
``modified effective LNRF parameter'' $\tilde q_\nr = M_0^{-1/2}
q_\nr$ and introduce functions $\tilde \gamma = \tilde q_\nr^2
\Gamma_0^2 {\cal DR}$ and $\tilde {\cal F}_\nr = \mathring{\cal F}
- \tilde \gamma \kappa_0^2$, we can write $\delta_0 \propto \tilde
{\cal F}_{\nr,0}$; thus, rays can reach the observer at $r = r_0$
with the desired frequency only if $\tilde {\cal F}_{\nr,0} \ge
0$. From the analysis of allowed regions in the $(r_0, b)$-plane
for LNRF observers (see Appendix D) we know that parameter $\tilde
q_\nr$ must be $< 1$ in order that there exists a non-empty class
of rays that arrive at the observer with the desired frequency,
and the rays must have $b$ from the interval(s) delimited by
functions $\tilde B_\pm = B_\pm(q \to \tilde q_\nr)$ (see equation
(\ref{eq:Bpm})), in order that they fall into that class. Which
rays are allowed depends on whether $r_0$ is greater or less than
radius $\tilde r_{0\a}$ at which function $\tilde B_-$ has an
asymptote: for $r_0 > \tilde r_{0\a}$ the rays must have $\tilde
B_- < b < \tilde B_+$, while for $r_0 < \tilde r_{0\a}$ they need
to have either $b < \tilde B_+$ or $b > \tilde B_-$. Thus,
similarly to the problem with LNRF observers, there are two
distinct allowed regions in the $(r_0, b)$ plane: region $\tilde
{\cal O}_\i$ extended over all $r_0$'s, with $\tilde B_- < b <
\tilde B_+$ for $r_0 > \tilde r_{0\a}$ and $b < \tilde B_+$ for
$r_0 < \tilde r_{0\a}$; and region $\tilde {\cal O}_\ii$ extended
over the interval $r_0 < \tilde r_{0\a}$ only, with $b > \tilde
B_-$.

\vskip 2mm At the boundaries of regions $\tilde {\cal O}_\i$ and
$\tilde {\cal O}_\ii$ rays do not have turning points as in the
problem with LNRF observers; instead, they have a nonzero rescaled
radial velocity $\eta_0$ there, $\eta_0 < 0$ at $\partial \tilde
{\cal O}_\i$ and $\eta_0 > 0$ at $\partial \tilde {\cal O}_\ii$.
Inside regions $\tilde {\cal O}_\i$ and $\tilde {\cal O}_\ii$ we
have two values of $\eta_0$ assigned to each point, one
($\eta_0^\bm$ in $\tilde {\cal O}_\i$ and $\eta_0^\bp$ in $\tilde
{\cal O}_\ii$) with the same sign as $\eta_0$ at the boundary of
the region and another ($\eta_0^\bp$ in $\tilde {\cal O}_\i$ and
$\eta_0^\bm$ in $\tilde {\cal O}_\ii$) with opposite sign in a
smaller region. In the latter case, the sign of $\eta_0$ changes
at lines $B_\pm$ given by equation ${\cal F}_{\nr,0} = 0$, where
${\cal F}_\nr = \mathring{\cal F} - \gamma \kappa_0^2$.

\vskip 2mm As $r_0$ decreases from $\infty$ to $r_\h$, $\Gamma_0$
increases from 1 to $\infty$, and as a result, $\tilde q_\nr$
increases for the given $q < 1$ from $q$ to 1 (since $\tilde q_\nr
= q_\nr/\big[1 + q_\nr^2 (1 - \Gamma_0^{-2}) \big]{}^{1/2} =
q/\big[(1 - q^2) \Gamma_0^{-2} + q^2\big]{}^{1/2}$). If the value
of $q$ is fixed, the form of regions $\tilde {\cal O}_\i$ and
$\tilde {\cal O}_\ii$ changes substantially in the vicinity of the
horizon in comparison with what we have obtained for LNRF
observers: rather than merge at $b = b_\h$ on the horizon, the
regions become separated by a gap between $b_{\h\pm} = b_\h/\big[1
\pm (1 - q^2)^{-1/2} qb_\h\big]$. For the parameters used in
Fig.~\ref{fig:allFobs}, $b_{\h+} = 1.045$ and $b_{\h-} = -2.188$.
(Thus, the gap lays entirely under the merging point at $b_\h =
4$.) If we want to preserve the form of regions $\tilde {\cal
O}_\i$ and $\tilde {\cal O}_\ii$ as they look in the problem with
LNRF observers, we need to fix $\tilde q_\nr$ rather than $q$; for
$r_0$ decreasing from $\infty$ to $r_\h$ we will then have $q$
decreasing from $\tilde q_\nr$ to 0 (since $q = \tilde
q_\nr/\big[(1 - \tilde q_\nr^2) \Gamma_0^2 + \tilde
q_\nr^2\big]{}^{1/2}$). Note that $q_\nr$ {\it increases} with
decreasing $r_0$ from $\tilde q_\nr$ to $\tilde q_\nr/(1 - \tilde
q_\nr^2)^{1/2}$, which is $> 1$ if $\tilde q_\nr
> 1/\sqrt{2}$. For such $\tilde q_\nr$ the constant $q_\nr$
surpasses 1 at some $r_{0,\l} > r_\h$, which means that functions
$B_\pm$ are defined only for $r_0 \ge r_{0,\l}$ -- there is no
change of sign of $\eta_0$ between $r_\h$ and $r_{0,\l}$. For the
value $\tilde q_\nr = 2/3$ used in Fig.~\ref{fig:allFobs},
functions $B_\pm$ are defined for all values of $r_0$.

\vskip 2mm We want to determine the lines dividing the $(r_0,
b)$-plane into regions ${\cal O}_{\h + \in}$, ${\cal O}_\h'$ and
${\cal O}_\in'$. For LNRF observers, we did it by computing
functions ${\cal B}_\pm$, equal for each $r_0$ to the value of $b$
at the lowest point of the upper forbidden corner and at the
highest point of the lower forbidden corner in the $(r, b)$-plane.
Calculation of ${\cal B}_\pm$ was straightforward, since equation
${\cal F}(r, r_0, b) = 0$ for the boundaries of the corners, lines
$b_\pm (r, r_0)$, was quadratic, and hence could be solved
analytically. Now we have two classes of rays with different
values of $\eta_0$ -- class I with $\eta_0 = \eta_0^\bp$ and class
II with $\eta_0 = \eta_0^\bm$, so we have also two functions $\cal
F$,
\begin{equation} \tag{E.2}
{\cal F}^\br = {\cal F}(\eta_0 = \eta_0^\br) = \mathring{\cal F} -
\tilde \gamma \big(M_0^{-1} N_0 \kappa_0^2 + \xi_0^2
\mathring{\cal F}_0 \mp 2\xi_0 M_0^{-1/2} \tilde {\cal
F}_{\nr,0}^{1/2} \kappa_0\big),
 \label{eq:Ff}
\end{equation}
where $N_0 = 1 - \gamma_0 \xi_0^2 = 1 - q_\nr^2 (1 -
\Gamma_0^{-2})$. Values of $b$ satisfying equations ${\cal F}^\br
= 0$ are now solutions to {\it quartic} equation ${\cal F}^\bp
{\cal F}^\bm = 0$. Even though there exists analytical solution in
this case, too, it is not too practical to determine functions
${\cal B}_\pm$ from it. Instead, we can compute ${\cal B}_\pm$ for
both $\eta_0^\bp$ and $\eta_0^\bm$ by looking for the values of
$b$ for which the minimum of ${\cal F}^\bp$ and ${\cal F}^\bm$
crosses zero (which may or may not happen for given $r_0$).

\vskip 2mm The lines we are searching for were, in the problem
with LNRF observers, just two parts of the boundary of region
${\cal O}_{\h + \in}$, separated  by points $P_\pm$. Now they are
still parts of that boundary, but they are separated by finite
segments of lines $\tilde B_\pm$ rather than by points lying on
them. At the endpoints of these segments, points $P_{A+}$ and
$P_{B+}$ on the line $\tilde B_+$ and points $P_{A-}$ and $P_{B-}$
on the line $\tilde B_-$, there are rays whose value of $b$ is
either maximum (in points $P_{A+}$, $P_{B+}$) or minimum (in
points $P_{A-}$, $P_{B-}$) among all rays coming to the observer
with the desired frequency. Rays in points $P_{A\pm}$ arrive from
and rays in points $P_{B\pm}$ head for the peak of the forbidden
region in the $(r, b)$-plane. Thus, radial coordinates of the
peaks, radii $R_{A\pm}$ and $R_{B\pm}$, differ from radial
coordinates of the observers, radii $R_{0A\pm}$ and $R_{0B\pm}$:
radii $R_{A\pm}$ are greater than $R_{0A\pm}$ and radii $R_{B\pm}$
are less than $R_{0B\pm}$. For the parameters used in
Fig.~\ref{fig:allFobs}, we have $(R_{0A+}, R_{A+}) = (1.708,
1.900)$, $(R_{0B+}, R_{B+}) = (2.405, 1.919)$, $(R_{0A-}, R_{A-})
= (2.859, 3.644)$ and $(R_{0B-}, R_{B-}) = (4.548, 3.624)$.

\vskip 2mm Line ${\cal B}_+^\bp$ lying on the border of the upper
part of region ${\cal O}_{\h + \in}^\bp$, adjacent to the arc
$P_{A+} P_{B+}$, line ${\cal B}_-^\bp$ forming the boundary of the
lower part region ${\cal O}_{\h + \in}^\bp$, adjacent to the arc
$P_{A+} P_{B+}$, as well as lines ${\cal B}_\pm^\bm$ demarcating
region ${\cal O}_{\h + \in}^\bm$, are all tangent to the
boundaries of region $\tilde {\cal O}_\i$, lines ${\cal B}_+^\br$
to $\tilde B_+$ and lines ${\cal B}_-^\br$ to $\tilde B_-$, just
as in the problem with LNRF observers lines ${\cal B}_\pm$ were
tangent to lines $B_\pm$. To see why, consider lines in the
vicinity of point $P_{B+}$. In the vertical band of the diagram
for rays of class I, where the boundary of region ${\cal O}_{\h +
\in}$ crosses the lines $r_0 = const$ three times, we have, in
addition to forbidden corner, a separate forbidden region bounded
from both sides in the form of a slant cigar. Rays with $b =
\tilde B_+$ on the right of the point $P_{B+}$, at $r_0 >
R_{0B+}$, bounce back from the lower edge of a forbidden {\it
corner} after passing by the observer, while those on the left of
the point $P_{B+}$, at $r_0 < R_{0B+}$, hit the upper edge of a
forbidden {\it cigar}. This leads, just as in the problem with
LNRF observers, to inequalities $dB_+/dr_0 \ge d{\cal
B_+}^\bm/dr_0$ for $r_0 \to R_{0B+}^+$ and $dB_+/dr_0 \le d{\cal
B}_+^\bp/dr_0$ for $r_0 \to R_{0B+}^-$. The difference is that now
for $r_0$ passing through $R_{0B+}$ from values $> R_{0B+}$,
forbidden corner first shrinks to a point and then stretches to a
forbidden cigar, while before, as $r_0$ passed through $R_{0+}$,
forbidden corner just shifted to the left without changing its
form. However, numerical calculation suggests that the
transformation from a corner to a cigar, even though it seems
discontinuous, is smooth enough to preserve the value of $d{\cal
B}/dr_0$.

\vskip 2mm In the problem with freely falling observers, regions
${\cal O}_{\h + \in}$ for rays of classes I and II look
significantly different, each in its own way, from region ${\cal
O}_{\h + \in}$ in the problem with LNRF observers. It is not only
that, as we just discussed, they are attached to forbidden regions
at finite segments of their boundaries rather than touching them
at a single point; region ${\cal O}_{\h + \in}$ also shrinks to a
point at a pair of points $Q_\pm$ lying on the lines $B_\pm$ for
rays of class I, and it is attached to the horizon in a finite
interval of values of $b$ for rays of class II.

\vskip 2mm As seen in Fig.~\ref{fig:allFobs}, the boundary of
${\cal O}_{\h + \in}^\bm$ consists of two parts, arcs $P_{A+}
P_{A-}$ and $P_{B+} P_{B-}$ intersecting at points $Q_\pm$. The
intersection at $Q_-$ has the form of a ``straight cross'', the
cross at $Q_+$ is ``inclined'' (the detailed numerics shows that
in fact the arc $P_{B+} P_{B-}$ at $Q_+$ is first deflected
slightly to the left and only then turns sharply to the right). In
the vicinity of point $Q_-$ there are two forbidden regions in the
$(r, b)$-plane, a cigar and an upper corner in a finite height
above it, and as $r_0$ passes through the radial coordinate of
point $Q_-$, $r_{Q-} = 3.494$, the cigar shrinks to a point and
then stretches again, leaning to the other side than before. To
the right of point $Q_+$, there is a cigar and a lower corner
which approach each other as $r_0$ passes through the radial
coordinate of point $Q_+$, $r_{Q+} = 2.159$, and at $r_0 < r_{Q+}$
only the corner remains. Since the left boundary of the central
part of region ${\cal O}_{\h + \in}^\bp$ turns to the right under
the point $Q_+$, minimum radii $(r_0, r)$ in that region are just
$\rho_{0A}^\bp = \rho_A^\bp = r_{Q+}$. For maximum radii numerical
calculation yields $(\rho_{0B}^\bp, \rho_B^\bp) = (3.966, 3.161)$.

\vskip 2mm The endpoints of the interval of $b$ where region
${\cal O}_{\h + \in}^\bm$ is attached to the horizon coincide with
impact parameters $\mathring{b}_{\ph\pm}$ of photon orbits in
vacuum. To see that, express additional term in $\cal F$ due to
the presence of plasma as $\Delta{\cal F} = - k_0^2 {\cal D}{\cal
R}$, where $k_0 = q_\nr \Gamma_0 \tilde \kappa_0$ and $\tilde
\kappa_0 = \kappa_0 - \xi_0 \eta_0$, write the factor $\tilde
\kappa_0$ as $\tilde \kappa_0 = \kappa_0 - \xi_0(- L_0 -
\sqrt{\delta_0})/M_0 = \big[\kappa_0 + \xi_0(M_0 \tilde {\cal
F}_{\nr,0})^{1/2}\big]/M_0$, insert into the expression under the
square root $ \tilde {\cal F}_{\nr,0} = \mathring{\cal F}_0 -
\tilde \gamma_0 \kappa_0^2 = (1 - \tilde q_\nr^2){\cal A}_0 -
{\cal D}_0 u_0^2 b^2/{\cal A}_0$ (this is just the expression for
${\cal F}_0$ obtained at the beginning of Appendix D, with $q$
replaced by $q_\nr$) and take the limit $r_0 \to r_\h$. It holds
$\xi_\h = -{\cal A}_\h^{-1/2}$, $M_\h = 1 + q_{\nr}^2 = 1/(1 -
\tilde q_{\nr,\h}^2)^{1/2}$ (since $q_{\nr,\h} = (q_0\Gamma_0)(r_0
\to r_\h) = \tilde q_\nr/(1 - \tilde q_\nr^2)^{1/2}$) and $\tilde
{\cal F}_{\nr,\h} = (1 - \tilde q_\nr^2){\cal A}_\h$, so that
$\tilde \kappa_\h \propto \kappa_\h - |\kappa_\h| = 0$ (since
$\kappa_\h > 0$ for $b = \mathring{b}_{\ph\pm}$). Thus, for
$\epsilon_0 = r_0 / r_\h \ll 1$ we have $\tilde \kappa_0 =
O(\epsilon_0)$, and since $\Gamma_0 = O(\epsilon_0^{-1/2})$, we
have also $k_0 = O(\epsilon_0^{1/2})$ and $k_\h = 0$. We see that
function ${\cal F}(r, r_0, b)$ reduces to $\mathring{\cal F}(r,
b)$ at $r_0 = r_\h$; therefore, functions ${\cal B}_\pm (r_0)$,
defined by equations ${\cal F}(r, r_0, b) =
\partial_r{\cal F}(r, r_0, b) = 0$, reduce to
$\mathring{b}_{\ph\pm}$ at $r_0 = r_\h$. For Kerr parameter $a =
0.8$ used in  Fig.~\ref{fig:allFobs}, this yields extreme values
of $b$ for rays seen by an observer on the horizon (in the sense
of limit) $({\cal B}_{\h+}, {\cal B}_{\h-}) = (3.237, -6.662)$.
Finally, since region ${\cal O}_{\h + \in}^\bm$ is adjacent to the
horizon, the radii from which radiation has access to the black
hole are bounded only from above: for the parameters used in
Fig.~\ref{fig:allFobs}, maximum radii $(r_0, r)$ are
$(\rho_{0B}^\bm, \rho_B^\bm) = (11.378, 3.067)$.

\enddocument